\documentclass[a4paper,12pt]{article}
\usepackage{amsmath,amsfonts,amssymb,amsthm,mathrsfs,graphicx,float,colordvi,url,wrapfig}
\usepackage{here,color,ulem,enumitem,bm,tikz-cd,blkarray,comment,slashed,cite}
\usepackage{bbm}
\usepackage[all]{xy}
\usepackage{tikz}  
\usepackage{scalerel}
  
\makeatletter

\newcommand{\xleftrightarrow}[2][]{\ext@arrow 3359\leftrightarrowfill@{#1}{#2}} 
\newcommand{\xdashrightarrow}[2][]{\ext@arrow 0359\rightarrowfill@@{#1}{#2}}
\newcommand{\xdashleftarrow}[2][]{\ext@arrow 3095\leftarrowfill@@{#1}{#2}}  
\newcommand{\xdashleftrightarrow}[2][]{\ext@arrow 3359\leftrightarrowfill@@{#1}{#2}} 
\def\rightarrowfill@@{\arrowfill@@\relax\relbar\rightarrow}
\def\leftarrowfill@@{\arrowfill@@\leftarrow\relbar\relax}
\def\leftrightarrowfill@@{\arrowfill@@\leftarrow\relbar\rightarrow}
\def\arrowfill@@#1#2#3#4{
$\m@th\thickmuskip0mu\medmuskip\thickmuskip\thinmuskip\thickmuskip
\relax#4#1 
\xleaders\hbox{$#4#2$}\hfill
#3$
}

\makeatother 
 
\def\p{{\partial}} 

\def\Tr{{\rm Tr}} 
\def\tr{{\rm tr}}

\def\ln{{\rm ln}} 

\def\Det{{\rm Det}}

\def\diag{{\rm diag}} 

\def\bdelta{\boldsymbol{\delta}}
\def\tg{\tilde{g}}

\def\slash#1{\not\!\!#1}

\def\p{\partial}

\begin{document}
\renewcommand{\thefootnote}{\fnsymbol{footnote}} 
\begin{titlepage}

\vspace*{6mm}

\begin{center}

{\large 
\textbf{Anomaly Equation of the Large U(1) Chiral Symmetry}
}

\vspace*{10.0mm} 
\normalsize
{\large Shingo Takeuchi}
\vspace*{1.5mm} 

%\textit{
%\small Institute of Research and Development, Duy Tan University, Da Nang, Vietnam}\\
%\vspace*{0.5 mm}
\textit{
\small Faculty of Environmental and Natural Sciences, Duy Tan University, Da Nang, Vietnam}\\
\end{center}

\vspace*{-1.0mm} 
\begin{abstract}
\vspace*{+0.0mm} 
In this study, we first heuristically construct the charges 
corresponding to the chiral transformation associated with the large U(1) gauge symmetry.
We refer to these as \textit{the large chiral charges}, 
and to the chiral transformation they generate as \textit{the large chiral transformations}.
Then, showing that 
these large chiral charges can be obtained based on Noether's theorem, 
we obtain the anomaly equation associated with these large chiral transformations.
%---

Subsequently, 
considering the one-loop diagrams of the fermionic field coupled to multiple classical gauge fields 
(which constitute the effective action of the model in this study with regard to the gauge field), 
we perform an axialization. 
%---
Then, defining the BRS transformations for the large U(1) gauge symmetry 
(we refer to these as \textit{the large BRS transformations}), 
we perform these transformations to these axialized one-loop diagrams, 
and demonstrate that the anomaly equations mentioned above can be derived 
by evaluating these diagrams.
We further confirm that these anomaly equations can be derived using the Fujikawa method.
%---

Finally, we discuss the breaking of unitarity
and the low-energy effective model
associated with the large chiral anomaly in this study,
and comment on potential future developments arising from them.

\end{abstract}
\end{titlepage}

\newpage  
  
\allowdisplaybreaks 
\setcounter{footnote}{0}

%=============================================
\section{Introduction}  
\label{zeqet}   
%=============================================   

The four-dimensional gravity 
described by the Einstein-Hilbert action includes 
unrenormalizable ultra-violet divergence. 
%--- 
In order to overcome this, 
extended gravitational theories have been proposed, 
such as
causal dynamical triangulation~\cite{Ambjorn:2012jv,Loll:2019rdj}, 
loop quantum gravity~\cite{Gambini:2011zz,Rovelli:2014ssa}, 
and superstring theory~\cite{Green:1987sp,Polchinski:1998rq,Hashimoto:2012vsa}.
%--- 
The asymptotic safety of gravitational theories has also been studied~\cite{Nagy:2012ef,Percacci:2017fkn}.

Meanwhile, 
the holographic principle holds~\cite{tHooft:1993dmi,Susskind:1994vu,Bigatti:1999dp,Bousso:2002ju}. 
%---
This suggests that
there is some microscopic structure in the gravity, 
which is considered to be encoded in one lower dimensional spacetime 
and to appear as gravity at low energy. 
%---
The holographic properties of gravity are currently under investigation, 
and one issue closely related to holography is the asymptotic symmetry group ({\it ASG}).

The ASG in hep-th is usually composed of general coordinate transformations
that preserves the fall-off conditions (usually denoted as ${\cal O}(r^{-p})$),
under which observationally indistinguishable configurations are identified.
%--- 
The ASG now usually investigated in hep-th are some ASG
on the asymptotic flat spacetime~\cite{Barnich:2009se,Strominger:2017zoo} 
or AdS$_3$~\cite{Brown:1986nw} 
(some ASG on dS is also considered~\cite{Anninos:2010zf}).
\newline

In the context of ASG on asymptotic flat spacetimes, 
an ADM mass~\cite{Regge:1974zd}
or infinite charges~\cite{Strominger:2017zoo} can be obtained  
depending on how the fall-off conditions are set.
%---
In the latter case, 
the ASG symmetries are broken 
(their central charges have been analyzed~\cite{Barnich:2010eb,Barnich:2011mi}), 
which 
is one of the issues
currently being investigated 
to solve the information paradox~\cite{Strominger:2017aeh}. 

Moreover, the latter case has been developed to various studies such as 
some equivalences between Ward identities and gravitational soft theorems~\cite{He:2014laa,
Campiglia:2015kxa,Laddha:2017ygw,Sahoo:2018lxl,AtulBhatkar:2018kfi}, 
holographic properties of the asymptotic flat spacetime~\cite{Raclariu:2021zjz,Pasterski:2021rjz}
and some equivalences between soft theorems and memory effects~\cite{
Pasterski:2015tva,Mao:2017wvx,Pate:2017vwa,Hamada:2017atr,Chu:2019ssw}.
%************************************************
% C
The observational studies concerning the memory effects in gravitational waves have been reviewed in~\cite{Mitman:2024uss}.
Soft-hairs in photon spheres and black hole shadows have also been  studied~\cite{Sarkar:2021djs,Lin:2022ksb}.

On the other hand, 
the Brown-Henneaux charge 
was obtained in the asymptotic AdS$_3$, 
which can be identified with the central charge of the Virasoro algebra of the CFT$_2$
on the boundary of the AdS$_3$~\cite{Brown:1986nw} (a derivation by the Fujikawa method was performed in~\cite{Terashima:2001gn}).
%---
Applying this to the near-horizon geometry of the D1-D5-P system,
the entropy microscopically obtained in~\cite{Callan:1996dv,Maldacena:1996ds} 
was reproduced~\cite{Strominger:1997eq}.
%---
Moreover, Kerr/CFT correspondence was discovered 
in the near-horizon geometry of the extremal Kerr black hole~\cite{Guica:2008mu}.
A correspondence analogous to Kerr/CFT was obtained
based on an explicit D-brane configuration given by the near-horizon geometry of the rotating extremal NS5-branes~\cite{Nakayama:2008kg}. 
%---

Exploiting the fact that flat spacetime and AdS are interchanged with each other 
up to the AdS radius,  
various relations between their thermodynamics, boundary CFT, etc are studied 
in connection with the ASG~\cite{Barnich:2012aw,Barnich:2012xq,Barnich:2012rz,Barnich:2013yka}. 
\newline

An analogous group to the ASG can also be considered in gauge theory, 
which is referred to as 
\textit{the large gauge group}. 
In a review article~\cite{Strominger:2017zoo}, 
large U(1) charges are introduced 
based on the U(1) charge   
by mixing some arbitrary function $\varepsilon$ 
in its expression.  
%---
These are defined for each choice of $\varepsilon$.
Consequently, an infinite set of large U(1) charges exists 
as $\varepsilon$ can take infinitely many forms.
The infinite transformations each charge generates form a large gauge symmetry group. 
The studies
\cite{ 
Campiglia:2015qka,Campiglia:2016hvg,Campiglia:2019wxe,AtulBhatkar:2019vcb,
AtulBhatkar:2020hqz,AtulBhatkar:2021sdr,Cheng:2022xyr,Banerjee:2021llh,
Fernandes:2023xim,Pasterski:2015zua,Baulieu:2024oql} 
can be regarded as investigations of large U(1) gauge symmetry.
As the study related with the present study, 
\cite{Baulieu:2024oql} discusses the BRST formalism 
including the large gauge symmetries and the regularization of the UV- and IR-divergences 
(due to the large gauge symmetries) in the Ward-Takahashi identity. % within the framework of that.

Now, we note that 
a chiral transformation can be considered 
when there is a gauge symmetry. 
Then, since there are large gauge symmetries, 
one can consider the chiral transformations associated with them.
However, this has not been investigated; 
therefore, we will address them in this study.

This study is expected to contribute to the understanding of chiral symmetries 
and their anomalies. 
%---
In addition, as anomalies are closely related to the breaking of gauge symmetry 
brought about by the regularization of UV divergence, 
this study is also expected 
to contribute to clarifying the connection 
between large gauge symmetries and these regularizations.
%---
Moreover, we can construct the low-energy effective model based on the large chiral anomalies in this study.
%---
We discuss these topics in  Sec.\,\ref{psent}.
\newline

We mention the organization of this paper.
The model addressed in this study is given in Sec.\,\ref{srhr}, 
and Noether's second theorem utilized in Sec.\,\ref{vhtew2} is reviewed in Sec.\,\ref{srebw}. 
%---
Then, once the U(1) charge is given on the Cartesian coordinates in Sec.\,\ref{etxga},
it is given again on the Penrose coordinates in Sec.\,\ref{zfhdsv}. 
The Penrose coordinates and their relationship with the Cartesian coordinates are reviewed
in Sec.\,\ref{efbg}.
%---
In Sec.\,\ref{vhtew1}, the large U(1) charges are given on the Penrose coordinates 
based on the U(1) charge given in Sec.\,\ref{zfhdsv}, 
then, in Sec.\,\ref{vhtew2}, these are given again on the Cartesian coordinates.

In Sec.\,\ref{rymntr}, the chiral charges are heuristically constructed
based on the large U(1) charges given in Sec.\,\ref{vhtew2}. 
We refer to the chiral transformations 
generated by these charges 
as \textit{the large chiral transformations}.
Then, it is shown that 
the large chiral charges heuristically constructed in Sec.\,\ref{rymntr} 
can be defined with Noether's theorem reviewed in Sec.\,\ref{srebw}. 
%---
In Sec.\,\ref{ykiwf}, 
the anomaly equation associated with the large chiral symmetry is heuristically obtained. 

In Sec.\,\ref{ypvsb},
defining the BRS transformations for the large U(1) gauge symmetry 
(which we refer to as the large BRS transformation), 
the one-loop diagrams 
obtained by axializing the one-loop diagrams of the fermionic field 
coupled to multiple classical gauge fields 
are considered 
(these one-loop diagrams constitute the effective action for the model in this study). 
Then, performing the large BRS transformations to these axialized one-loop diagrams,
it is shown that 
the anomaly equations obtained in Sec.\,\ref{ykiwf} can be derived. 

In Sec.\,\ref{itioh}, it is shown that 
the anomaly equation obtained in Sec.\,\ref{ykiwf} and \ref{ypvsb}
can be derived using the Fujikawa method. 
In Sec.\,\ref{psent}, issues related to the anomaly of the large chiral symmetry in this study, 
and a possible future development based on them are discussed.
In Sec.\,\ref{iptr}, this study is summarized.

In Appendix\,\ref{app:ttih}, the derivation of some equations in Sec.\,\ref{srebw} is noted, 
and, in Appendix\,\ref{app:orrew}, the formulas used in Sec.\,\ref{ypvsb} are noted.
In Appendix\,\ref{app:stod}, the Ward-Takahashi identity and the Wess-Zumino consistency condition 
are derived for use in the discussion in  Sec.\,\ref{psent1}.

\vspace{-1.0mm}
%=============================================
\section{The study model} 
\label{srhr}
%============================================= 
 
The model considered in this study is given as follows:
\vspace{-1.0mm}
\begin{eqnarray}\label{erer}
S_0= \int d^4x \,\sqrt{-g} \, {\cal L}_0, \quad 
{\cal L}_0=
-\frac{1}{4}F_{\mu\nu}F^{\mu\nu}
+\bar{\psi} i \! \slash{D}\psi, 
\end{eqnarray}
where $\slash{D}\equiv\gamma^\mu D_\mu$ ($D_\mu\equiv\partial_\mu-ieA_\mu$), 
$\bar{\psi}\equiv\psi^\dagger \gamma_0$,  
$g$ means the determinant of the metric and  
$A_\mu$ is the U(1) gauge field. Thus, $F_{\mu\nu}=\partial_\mu A_\nu-\partial_\nu A_\mu$. 
$\psi$ is the Dirac field, which is massless in this study. 

\vspace{-1.0mm}
%=============================================
\section{Noether's second theorem}
\label{srebw}
%============================================= 

In this study, we address the charges 
in the case that transformation parameters depend on coordinates. 
%---
Therefore, we give the general expression of charges in Noether's second theorem. 
%---
The results in this section are utilized 
in Sec.\,\ref{zfhdsv} and \ref{vhtew1}.
%---

\vspace{-1.0mm}
%=============================================
\subsection{Noether's first theorem}
\label{app:stod1}
%=============================================

We first discuss Noether's first theorem as it provides the necessary background for explaining Noether's second theorem. 
%---
Considering a four-dimensional spacetime $\Omega$ 
patched by the coordinates $x^\mu$, 
we write a given Lagrangian as $L(\phi_A,\phi_{A,\mu})$, 
where $\phi_A=\phi_A(x)$ represent all fields ($A =1,\cdots, N$ identify each field) 
and $\phi_{A,\mu} \equiv \partial_\mu\phi_A(x)$. 
Then, considering the action: $I =\int_\Omega d^4x \, L$,
the Euler-Lagrange ({\it EL}) equation for each $\phi_A$ can be given as 
\vspace{-1.0mm}
\begin{eqnarray}\label{wevawr}
[L]^A \equiv \frac{\partial L}{\partial \phi_A}-\partial_\mu (\frac{\partial L}{\partial \phi_{A,\mu}}).  
\end{eqnarray}
Next, we write the constant transformations (global transformations) as follows:
\vspace{-1.0mm}
\begin{eqnarray}\label{redz}
\delta x^\mu  \equiv \sum_{r=1}^n \eta^r X^\mu{}_r, \quad
\delta \phi_A \equiv \sum_{r=1}^n \eta^r M_{r,A}, 
\end{eqnarray}
where $\eta^r$ ($r=1,2, \cdots,n$) represent the constant transformation parameters. 
$X^\mu{}_r$ and $M_{r,A}$ are functions with regard to $x^\mu$ and $\phi_A$ 
($M_{r,A}$ may generally depend on the coordinates). 
Then, the variation of $I$ under (\ref{redz}) can be written as $\int d^4x \, \eta^r \, \partial_\mu(\cdots)$. 
From this, the following conservation law can be obtained:
\vspace{-1.0mm}
\begin{eqnarray}\label{vrtyz}
\partial_\mu J^\mu_r = 0, 
\quad
%---
J^\mu_r \equiv
\frac{\partial L}{\partial \phi_{A,\mu}}M_{r,A}
-T^\mu{}_\nu \, X^\nu{}_r, 
\quad 
%---
T^\mu{}_\nu \equiv 
\frac{\partial L}{\partial \phi_{A,\mu}}\phi_{A,\nu}-\delta^\mu{}_\nu L
\end{eqnarray}
for each $r$. 
From (\ref{vrtyz}), one charge for each $r$ can be defined as
\vspace{-1.0mm} 
\begin{eqnarray}\label{trshk}
Q_r^{\rm (G)} \equiv \int_\Sigma \, d^3x \, J^0_r,
\end{eqnarray}
where $\Sigma$ means a $x^0$-constant 3D hypersurface on $\Omega$. 

\vspace{-1.0mm}
%=============================================
\subsection{Noether's second theorem}
\label{app:ssdik}
%=============================================

Now, we write the local transformations as follows:
\vspace{-1.0mm}
\begin{eqnarray}\label{gdwr}
\delta x^\mu  \equiv \lambda^r X^\mu{}_r, \quad  
\delta \phi_A \equiv \lambda^r M_{r,A} + \partial_\mu \lambda^r N_r{}^\mu{}_{,A},
\end{eqnarray}
where
$\lambda^r =\lambda^r (x)$ are the transformation parameters. 
$X^\mu{}_r$, $M_{r,A}$ and $N_r{}^\mu{}_{,A}$ are functions 
of $x^\mu$ and $\phi_A$. 
We use the same notations as those in (\ref{redz}).  
%---
If $I$ is invariant under (\ref{gdwr}), 
the following equations hold: 
\vspace{-1.0mm}
\begin{eqnarray}\label{resyu}
\partial_\mu B^\mu{}_r =0, \quad 
B^\mu{}_r +\partial_\nu C^{\nu \mu}{}_r =0, \quad 
C^{\mu\nu}{}_r+C^{\nu\mu}{}_r =0,
\\*[1.0mm]
B^\mu{}_r     
\equiv
J^\mu_r 
+[L]^A N_r{}^\mu{}_{,A}, \quad
C^{\mu\nu}{}_r
\equiv\frac{\partial L}{\partial \phi_{A,\mu}} N_r{}^\nu{}_{,A}, 
\nonumber
\end{eqnarray}
where $J^\mu_r$ is the same as $J^\mu_r$ in (\ref{vrtyz}). 
The derivation of these is given in Appendix\,\ref{app:ttih}.
From the first equation in (\ref{resyu}), 
a conservation law can be obtained for each $r$ as follows:
\vspace{-1.0mm}
\begin{eqnarray}\label{rkta}
\partial_\mu \, J^\mu_r =0, 
\end{eqnarray}
where the EL equations (\ref{wevawr}) have been utilized.  
Therefore, for each $r$, one charge can be defined for example
\vspace{-1.0mm}
\begin{eqnarray}\label{rrwk}
Q_r^{\rm (L)} \equiv
\int_\Sigma \, d^3x \,J^0_r, 
\end{eqnarray}
where $\Sigma$ means a $x^0$-constant hypersurface as well as $\Sigma$ in (\ref{trshk}).

\vspace{-1.0mm}
%=============================================
\section{The U(1) charge on the Cartesian coordinates}
\label{etxga}
%============================================= 

In the previous section, we generally defined the charge with regard to the local transformation.
Using that definition, we present the U(1) charge in our model on the Cartesian coordinates. 
\newline

The equations of motion with regard to $A_\mu$, $\psi$ and $\bar{\psi}$ can be obtained 
from (\ref{erer}) as follows:
\vspace{-1.0mm}
\begin{eqnarray}\label{urst}
\nabla_\nu F^{\nu\mu}-J_{\rm U(1)}^\mu=0, \quad 
i\!\slash{D}\psi =0, \quad
iD_\mu \bar{\psi} \gamma^\mu =0,
\quad 
J_{\rm U(1)}^\mu \equiv -e\,\bar{\psi} \gamma^\mu \psi.
\end{eqnarray}
The Lagrangian density $\sqrt{-g}\,{\cal L}_0$ in (\ref{erer}) is invariant 
under the following U(1) gauge transformation: 
\vspace{-1.0mm}
\begin{subequations}\label{vwrer}
\begin{align} 
\label{vwrer1}
& 
A_\mu    \to \,\, A_\mu+ e^{-1}\,\partial_\mu \lambda, \quad
\\*[1.0mm]
\label{vwrer2}
& 
\psi     \to \,\, e^{ie\lambda}  \, \psi,      \quad
\bar{\psi} \to      e^{-ie\lambda} \, \bar{\psi},
\end{align} 
\end{subequations}
where $\lambda=\lambda(x)$. 
%---
Taking $\lambda$ to the linear order, we denote (\ref{vwrer}) by the notations (\ref{gdwr}). 
Then, $X^\mu{}_r$, $M_{r,A}$ and $N_r{}^\mu{}_{,A}$ in (\ref{gdwr}) are taken as 
$X^\mu=0$ and
\vspace{-1.0mm}
\begin{subequations}\label{ssrr}
\begin{align} 
\label{ssrr2}
(M_{,0}, M_{,1}, M_{,2}, M_{,3}, M_{,4}, M_{,5})
=& \,\, (0,0,0,0,ie\psi,-ie\bar{\psi}),
\\*[1.0mm]
\label{ssrr3}
(N^\mu{}_{,0}, N^\mu{}_{,1}, N^\mu{}_{,2}, N^\mu{}_{,3}, N^\mu{}_{,4}, N^\mu{}_{,5})
=& \,\, e^{-1}(\delta^\mu{}_0, \delta^\mu{}_1, \delta^\mu{}_2, \delta^\mu{}_3, 0, 0),
\end{align} 
\end{subequations}
where $\mu=0,\cdots,3$. 
We have taken $\phi_A$ as $(\phi_0,\phi_1,\phi_2,\phi_3,\phi_4,\phi_5)=(A_0,A_1,A_2,A_3,\psi,\bar{\psi})$. 
We have omitted to denote $r$ 
as the index $n$ is $1$, which is $r_{\rm max}$.
Applying (\ref{ssrr}) to (\ref{resyu}),  
\vspace{-1.0mm}
\begin{eqnarray}\label{j5ir}
B^\mu 
=  \frac{\partial (\sqrt{-g}\,{\cal L}_0)}{\partial (\partial_\mu \psi)}(ie \psi)
  +\frac{\partial (\sqrt{-g}\,{\cal L}_0)}{\partial (\partial_\mu \bar{\psi})}(-ie \bar{\psi})
=  \sqrt{-g} \, J_{\rm U(1)}^\mu, \quad \!\!\!\!
 C^{\mu\nu} 
=  \frac{\partial (\sqrt{-g}\,{\cal L}_0)}{\partial (\partial_\mu A_\lambda)} e^{-1} \delta^\nu{}_\lambda
= -\sqrt{-g}\,F^{\mu\nu},
\end{eqnarray}
where the EL equations (\ref{wevawr}) have been utilized.
Then, the conservation law for the U(1) gauge transformation (\ref{vwrer}) can be obtained 
from the first one of (\ref{resyu}) as 
\vspace{-1.0mm}
\begin{eqnarray}\label{udvre}
0= \nabla_\mu (\sqrt{-g} \,J_{\rm U(1)}^\mu).
\end{eqnarray}

Now, we consider the following Cartesian coordinates:
\vspace{-1.0mm} 
\begin{eqnarray}\label{rmtrt}
ds^2 = (dx^0)^2 -(dx^1)^2 -(dx^2)^2-(dx^3)^2.
\end{eqnarray}
Then, following (\ref{rrwk}), 
the U(1) charge on the Cartesian coordinates can be given as
\vspace{-1.0mm}
\begin{eqnarray}\label{bebas}
Q_{\rm U(1)}^{\rm (M)} \equiv                  \int_\Sigma \, d^3x      \, J^0_{\rm U(1)}
                        =\,  \sum_{k=1}^3\,\int_S  \,    d\sigma_k \, F^{0k},
\end{eqnarray} 
where $\Sigma$ is a $x^0$-constant 3D hypersurface, 
and $S$ means its boundary.  

\vspace{-1.0mm}
%=============================================
\section{The Penrose coordinates}
\label{efbg}
%============================================= 

In this study, the chiral anomaly of the large U(1) gauge symmetry is addressed. 
We refer to it as \textit{the large chiral anomaly}.
%---
We define it on the Cartesian coordinates
by following the custom that the U(1) chiral anomaly is defined on these coordinates.  
However, the asymptotic symmetry is defined on the Penrose coordinates given with $(z, \bar{z})$ 
%---
(the reason to consider the $(z, \bar{z})$ coordinates is that, by doing that, 
it becomes easier to address ``the matching condition''~\cite{Strominger:2017zoo}).   
%---
Therefore, we review the link between the Cartesian coordinates and the Penrose coordinates, 
which is utilized in the following sections.
\newline

\begin{figure}[H]  
\vspace{-10.0mm} 
\begin{center}
\includegraphics[clip,width=8.5cm,angle=-90]{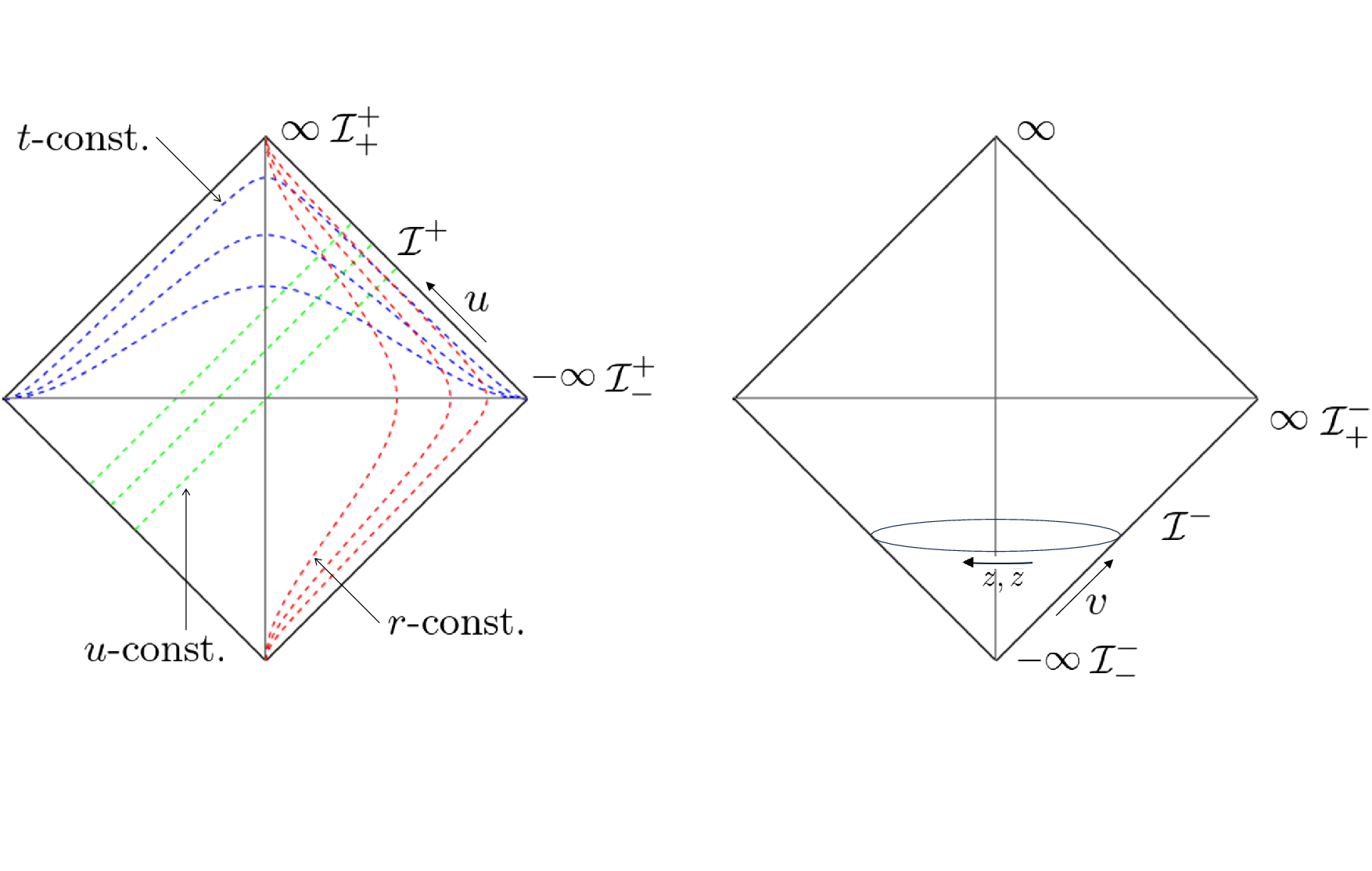} 
\end{center}
\vspace{-20.0mm}
\caption{The 4D flat spacetime with the coordinates (\ref{rntaw})
} 
\label{wsdd57}
\end{figure}

We consider the following coordinate transformation:
\vspace{-1.0mm}  
\begin{subequations}\label{gwgr}
\begin{align} 
x^0=& \, u + r,\quad
x^1=+ r \,\frac{1-z\bar{z}}{1+z\bar{z}},\quad 
x^2+ix^3= + \frac{2r z}{1+z\bar{z}} \quad \textrm{for $t \ge 0$},\\*[1.0mm]
x^0=& \, v - r,\quad
x^1=- r \,\frac{1-z\bar{z}}{1+z\bar{z}},\quad 
x^2+ix^3= - \frac{2r z}{1+z\bar{z}} \quad \textrm{for $t \le 0$},
\end{align} 
\end{subequations}
where $x^\mu$ are the Cartesian coordinates defined in (\ref{rmtrt}). 
$r^2 \equiv (x^1)^2+(x^2)^2+(x^3)^2$, 
and $(z, \bar{z})$ are the coordinates 
on a complex plane 
defined  as the destinations of the mappings of the points 
on the surface of the spatial $S^2$ in (\ref{rmtrt}). 
Using (\ref{gwgr}), $ds^2$ in (\ref{rmtrt}) can be given as 
\vspace{-1.0mm}
\begin{align}\label{rntaw}
ds^2 = 
\left\{ 
\begin{array}{ll}
du^2 +2du dr -2r^2 \,\gamma_{z\bar{z}}\, dz d\bar{z}  & \,\textrm{for $t \ge 0$}, \\*[1.0mm] 
dv^2 -2dv dr -2r^2 \,\gamma_{z\bar{z}}\, dz d\bar{z}  & \,\textrm{for $t \le 0$},
\end{array} 
\right.
\end{align}
where $\gamma_{z\bar{z}} \equiv \frac{2}{(1+z\bar{z})^2}$.
The 4D flat spacetime with the coordinates (\ref{rntaw}) can be seen in Fig.\,\ref{wsdd57}.
%==================
%===== FIGURE =====
%==================

We note the points in the Penrose coordinates
which are utilized in the next sections:
\vspace{-1.0mm}
\begin{itemize}
\item[1)]
The $x^0$-constant hypersurfaces infinitely exist for each $x^0$.
However, their boundaries on the $x^0 \ge 0$ and $x^0 \le 0$ regions 
are commonly given by ${\cal I}^+_-$ and ${\cal I}^-_+$. 

\item[2)]
On the $x^0 \ge 0$ and $x^0 \le 0$ region respectively,
the $x^0 \to \infty$ and $x^0 \to -\infty$ hypersurface overlaps the $r \to \infty$ hypersurface,
except for (${\cal I}^+_+$, ${\cal I}^+_-$) and (${\cal I}^-_-$, ${\cal I}^+_-$).  
\end{itemize}

\vspace{-1.0mm}
%=============================================
\section{The U(1) charge on the Penrose coordinates}
\label{zfhdsv}
%============================================= 

The large U(1) charges are defined by some extension of the U(1) charge on the Penrose coordinates~\cite{Strominger:2017zoo}. 
However, we gave the U(1) charge on the Cartesian coordinates in Sec.\,\ref{etxga}. 
Therefore, we transform its coordinates to the Penrose coordinates. 
\newline
 
First, we consider the coordinates $y^a \equiv (t,r,z,\bar{z})$, 
where $(z,\bar{z})$ and $r$ are those in (\ref{gwgr}).
Meanwhile, we denote the Cartesian coordinates (\ref{rmtrt}) as $x^\mu \equiv (t,\, x^1,x^2,x^3)$. 
$\tilde{J}_{\rm U(1)}^a$ and $\tilde{F}^{ab}$, quantities on the $y^a$ coordinates, are given 
from these on the $x^\mu$ coordinates as 
\vspace{-1.0mm}
\begin{eqnarray}\label{eerts}
\tilde{J}_{\rm U(1)}^a=h^a{}_\mu J_{\rm U(1)}^\mu, \quad
\tilde{F}^{ab}=h^a{}_\mu h^b{}_\nu F^{\mu\nu},\quad
h^a{}_\mu \equiv {\partial y^a}/{\partial x^\mu},
\end{eqnarray} 
where $J_{\rm U(1)}^m$ is defined in (\ref{urst}). 
With these, the U(1) charge (\ref{bebas}) on the coordinates $y^a$ are given as 
\vspace{-1.0mm}
\begin{eqnarray}\label{weweu}
\tilde{Q}_{\rm U(1)}
\equiv \int_\Sigma  \,    dr   \,   d^2z  \,\, r^2 \,\gamma_{z\bar{z}} \, \tilde{J}_{\rm U(1)}^t  
     = \int_S   \,        d^2z \,\,            r^2 \,\gamma_{z\bar{z}} \, \tilde{F}^{rt}.
\end{eqnarray} 

We change the coordinates of $\tilde{Q}_{\rm U(1)}$ to the Penrose coordinates (\ref{rntaw}).
We denote the Penrose coordinates $(u,r,z,\bar{z})$ and $(v,r,z,\bar{z})$ with the same $w^\mu$.
%---
$\widetilde{F}^{\mu\nu}$, quantities on the $w^m$ coordinates, 
are given from those on the $y^a$ coordinates as 
\vspace{-1.0mm}
\begin{eqnarray}\label{sekts}
\widetilde{J}_{\rm U(1)}^\mu=h^\mu{}_a \tilde{J}_{\rm U(1)}^a, \quad
\widetilde{F}^{\mu\nu}=h^\mu{}_a h^\nu{}_b \tilde{F}^{ab}, \quad
h^\mu{}_a\equiv {\partial w^\mu}/{\partial y^a}
\end{eqnarray} 
(we use the same notation ``$h$'' as (\ref{eerts}); 
we tell the difference by the indices).
With these, the U(1) charge (\ref{weweu}) on the $w^m$ coordinates  are given as 
\vspace{-1.0mm}
\begin{eqnarray}\label{thrsr}
\widetilde{Q}^{+} \equiv \int_{{\cal I}^+_-} \! d^2z \,r^2\,\gamma_{z\bar{z}}\, \widetilde{F}_{ru}, \quad 
\widetilde{Q}^{-} \equiv \int_{{\cal I}^-_+} \! d^2z \,r^2\,\gamma_{z\bar{z}}\, \widetilde{F}_{rv}.
\end{eqnarray}
$\widetilde{Q}^{+}$ and $\widetilde{Q}^{-}$ are those 
on the $x^0 \ge 0$ and $x^0 \le 0$ regions 
given by $(u,r,z,\bar{z})$ and $(v,r,z,\bar{z})$, respectively. 
%---
The boundaries of $t$-constant 3D hypersurfaces are always ${\cal I}^+_-$ and ${\cal I}^-_+$ 
on the $w^m$ coordinates, respectively, as can be seen in Fig.\,\ref{wsdd57}.
In addition, it is obvious that $\widetilde{Q}^{\pm}$ agree with each other 
as the integral value for a common $x^0$-constant surface at $x^0=0$.
These guarantee $\widetilde{Q}^{\pm}$ are conserved quantities. 

\vspace{-1.0mm}
%=============================================
\section{The large U(1) charges on the Penrose coordinates}
\label{vhtew1} 
%============================================= 

In the previous section, the U(1) charge was presented on the Penrose coordinates.
In this section, the large U(1) charges (\textit{the large charges}) are presented by extending that. 
\newline

Following \cite{Strominger:2017zoo}, the large gauge charges are defined 
by mixing $\varepsilon=\varepsilon(z,\bar{z})$ into $\widetilde{Q}^{\pm}$ in (\ref{thrsr}) as 
\vspace{-1.0mm}
\begin{eqnarray}\label{avkr}
\widetilde{Q}^{+}_\varepsilon \equiv \int_{{\cal I}^+_-} d^2z \,r^2\,\gamma_{z\bar{z}}\,\varepsilon\, \widetilde{F}_{ru}, \quad 
\widetilde{Q}^{-}_\varepsilon \equiv \int_{{\cal I}^-_+} d^2z \,r^2\,\gamma_{z\bar{z}}\,\varepsilon\, \widetilde{F}_{rv},
\end{eqnarray}
where $\varepsilon$ is an arbitrary function, which amplifies/weakens the electro-magnetic field for each angle.
%---
Since the function to be considered as $\varepsilon$ infinitely exists,
the value of $\widetilde{Q}^\pm_\varepsilon$ infinitely exists as well, 
and each of which is respectively constant against $x^0$ 
(this can be explained in the same way under (\ref{thrsr})).  
In what follows, we give $\widetilde{F}_{ru}$ and $\widetilde{F}_{rv}$ in (\ref{avkr}) in terms of the conserved current. 
\newline

From the Maxwell equations (\ref{urst}), 
$\nabla^\mu \widetilde{F}_{\mu u}=\widetilde{J}_{\rm U(1)}{}_u$ and 
$\nabla^\mu \widetilde{F}_{\mu v}=\widetilde{J}_{\rm U(1)}{}_v$. 
The metrices on the Penrose coordinates (\ref{rntaw}) are given as
\vspace{-1.0mm}
\begin{subequations}\label{erdsv}
\begin{align} 
& 
g_{uu}=+g_{ur}=+g_{ru}=1, 
\quad g_{z\bar{z}}=g_{\bar{z}z}=-r^2\gamma_{z\bar{z}} 
\quad \textrm{for $t \ge 0$}, 
\\*[1.0mm] 
& 
g_{vv}=-g_{vr}=-g_{rv}=1, 
\quad \hspace{0.5mm} g_{z\bar{z}}=g_{\bar{z}z}=-r^2\gamma_{z\bar{z}}
\quad \textrm{for $t \le 0$} 
\end{align} 
\end{subequations}
(others are 0, and $\gamma_{z\bar{z}}$ is given in (\ref{rntaw})).
From these, 
\vspace{-1.0mm}
\begin{subequations}\label{dsyk}
\begin{align} 
\label{dsyk1}
+\,\partial_u \widetilde{F}^{}_{ru}
=& \,\,
J^{}_{\rm U(1)}{}_u
-g^{z\bar{z}}\nabla_{\bar{z}}\widetilde{F}^{}_{zu}
-g^{\bar{z}z}\nabla_{z}\widetilde{F}^{}_{\bar{z}u},\\*[1.0mm]
\label{dsyk2}
-\,\partial_v \widetilde{F}^{}_{rv}
=& \,\,
J^{}_{\rm U(1)}{}_v
-g^{z\bar{z}}\nabla_{\bar{z}}\widetilde{F}^{}_{zv}
-g^{\bar{z}z}\nabla_{z}\widetilde{F}^{}_{\bar{z}v}.
\end{align} 
\end{subequations}
From (\ref{dsyk}), $\widetilde{F}_{ru}$ on ${\cal I}^+_-$ and $\widetilde{F}_{rv}$  on ${\cal I}^-_+$
can be given as follows:
\vspace{-1.0mm}
\begin{subequations}\label{ykssd}
\begin{align} 
\label{ykssd1}
\widetilde{F}_{ru}\vert_{{\cal I}^+_-}
=& \,\,
\widetilde{F}_{ru}\vert_{u=u_0}
+\int_{u_0}^{-\infty} \! du\, (
\widetilde{J}^{}_{\rm U(1)}{}_u
-g^{z\bar{z}}\nabla_{\bar{z}}\widetilde{F}^{}_{zu}
-g^{\bar{z}z}\nabla_{z}\widetilde{F}^{}_{\bar{z}u}),
\\*[1.0mm]
\label{ykssd2}
\widetilde{F}_{rv}\vert_{{\cal I}^-_+}
=& \,\,
\widetilde{F}_{rv}\vert_{v=v_0}
-\int_{v_0}^{\infty} \,\,dv\, (
\widetilde{J}^{}_{\rm U(1)}{}_v
-g^{z\bar{z}}\nabla_{\bar{z}}\widetilde{F}^{}_{zv}
-g^{\bar{z}z}\nabla_{z}\widetilde{F}^{}_{\bar{z}v}),
\end{align} 
\end{subequations}
where ${\cal I}^+_-$ and ${\cal I}^-_+$ can be reached 
by taking $u \to -\infty$ and $v \to +\infty$ on the infinite $r$ line;
therefore, $r$ is being taken to $\infty$ in (\ref{ykssd1}) and (\ref{ykssd2}).

When the values of $u_0$ and $v_0$ are finite, 
the values of $\widetilde{F}_{ru}\vert_{u=u_0}$ and $\widetilde{F}_{rv}\vert_{v=v_0}$ are concretely unclear, 
and (\ref{ykssd1}) and (\ref{ykssd2}) do not concretely make sense. 
However, 
the point to be reached by $u \to +\infty$ ($v \to -\infty$)
on the infinite $r$ line is ${\cal I}^+_+$ (${\cal I}^-_-$). 
In light of this, 
we take $u_0$  
to $+\infty$ ($v_0$ to $-\infty$).
Then, with the fact that $r$ in (\ref{ykssd}) is being taken to $\infty$ (as mentioned above), 
$\widetilde{F}_{ru}\vert_{u = u_0}$ and $\widetilde{F}_{rv}\vert_{v = v_0}$ 
are reduced to $\widetilde{F}_{ru}\vert_{{\cal I}^+_+}$ and $\widetilde{F}_{rv}\vert_{{\cal I}^-_-}$, respectively.
The values of $\widetilde{F}_{ru}\vert_{{\cal I}^+_+}$ and $\widetilde{F}_{rv}\vert_{{\cal I}^-_-}$ can be known as
\vspace{-1.0mm}
\begin{eqnarray}\label{nkvr}
\widetilde{F}_{ru}\vert_{{\cal I}^+_+}=\widetilde{F}_{rv}\vert_{{\cal I}^-_-} = 0
\end{eqnarray}
due to the facts and assumptions:
the electro-magnetic field never reaches ${\cal I}^+_+$ and ${\cal I}^-_-$, 
and no charge exists at ${\cal I}^+_+$ and ${\cal I}^-_-$ (as the Dirac field in this study is massless).
 
As such, taking $u_0$ and $v_0$ to $+\infty$ and $-\infty$, respectively, 
we substitute (\ref{ykssd1}) and (\ref{ykssd2}) in $\widetilde{Q}_{\varepsilon}^{\pm}$ in (\ref{avkr}).
As a result, the large gauge charges on the Penrose coordinates are finally given as
\vspace{-1.0mm}
\begin{subequations}\label{vrse1}
\begin{align} 
\label{vrse11}
\widetilde{Q}_{\varepsilon}^+  
=& 
\int_{\infty}^{-\infty}\!\! du \int_{{\cal I}^+_-} \! d^2z \, \varepsilon \, 
\Big\{
\!- (
  \nabla_z         \widetilde{F}_{\bar{z}u}
+ \nabla_{\bar{z}} \widetilde{F}_{zu})
+ 
\,\gamma_{z\bar{z}}\,\widetilde{J}_{\rm U(1)}{}_u
\Big\}, 
\\*[0.5mm]
\label{vrse12}
\widetilde{Q}_{\varepsilon}^-  
=& 
\int_{-\infty}^{\infty}  dv \int_{{\cal I}^-_+} d^2z \, \varepsilon \, 
\Big\{
\!+(
  \nabla_z         \widetilde{F}_{\bar{z}v}
+ \nabla_{\bar{z}} \widetilde{F}_{zv})
+ 
\, \gamma_{z\bar{z}}\,\widetilde{J}_{\rm U(1)}{}_v
\Big\}.
\end{align} 
\end{subequations}
We comment on the integral regions of (\ref{vrse11}) and (\ref{vrse12}). 
$r$ is taken to infinity in (\ref{vrse11}); 
therefore, by the property 2 in Sec.\ref{efbg}, 
the integral region in (\ref{vrse11}) overlaps with
the $t$-constant 3D hypersurface at $t \to +\infty$ 
except for ${\cal I}^+_+$ and ${\cal I}^+_-$ 
(these are the points given by $u\to \infty$ and $-\infty$, respectively).
%---
The mismatching at ${\cal I}^+_+$ can be ignorable by the assumptions under (\ref{nkvr}).
The mismatching at ${\cal I}^+_-$ can be ignorable by the fact that there is no electro-magnetic field at ${\cal I}^+_-$ 
if there is no charges at ${\cal I}^-_-$, and the assumption that  there is no charges at  ${\cal I}^+_-$.
%---
Therefore, the integral region of (\ref{vrse11}) can be regarded as a $t$-constant 3D hypersurface.
The integral region of (\ref{vrse12}) is considered in the same way.

\vspace{-1.0mm} 
%=============================================
\section{The large U(1) charges on the Cartesian coordinates}
\label{vhtew2}
%============================================= 

In the previous section, the large U(1) charges were presented on the Penrose coordinates. 
In this section, we transform their coordinates to the Cartesian coordinates $x^\mu$. 
\newline

With (\ref{eerts}) and (\ref{sekts}), 
$\widetilde{F}_{\bar{z}u}$, $\widetilde{F}_{\bar{z}v}$ and $\widetilde{J}_{\rm U(1)}{}_u$ 
can be given by those on the $x^\mu$ coordinates as  
\vspace{-1.0mm}
\begin{subequations}\label{reefg}
\begin{align}
%-----
\label{reefg1}
\widetilde{F}_{\bar{z}u}
=& \,\,\, h_{\bar{z}}{}^\mu F_{\mu t},\quad
%-----
\widetilde{F}_{zu}
= \, h_{z}{}^\mu F_{\mu t},\quad
%-----
\widetilde{J}^{}_{\rm U(1)}{}_u
= \, h_u{}^\mu J^{}_{\rm U(1)}{}_\mu.
\\*[1.0mm]
%-----
\label{reefg1}
\widetilde{F}_{\bar{z}v}
=& \,\,\, h_{\bar{z}}{}^\mu F_{\mu t},\quad
%-----
\widetilde{F}_{zv}
= \, h_{z}{}^\mu F_{\mu t},\quad
%-----
\widetilde{J}^{}_{\rm U(1)}{}_v
= \, h_v{}^\mu J^{}_{\rm U(1)}{}_\mu,
\end{align}
\end{subequations}
where $J^{}_{\rm U(1)}{}_\mu$ is defined in (\ref{urst}). 
Substituting these in (\ref{vrse1}),
the large gauge charges $\widetilde{Q}_{\varepsilon}^\pm$ in (\ref{vrse1}) 
can be commonly given on the $x^\mu$ coordinates as
\vspace{-1.0mm}
\begin{eqnarray}\label{dtre}
Q_{\varepsilon} 
\equiv
\int_\Sigma \, d^3x \,\varepsilon \, 
\Big(
  h_z{}^\mu \,\nabla_\mu         (h_{\bar{z}}{}^\nu F_{\nu  t})
+ h_{\bar{z}}{}^\mu \,\nabla_\mu (h_{z}{}^\nu F_{\nu  t})
\Big)
+ \int_\Sigma \, d^3x \, \varepsilon\, J^{}_{\rm U(1)}{}_t,
\end{eqnarray}
$\Sigma$ means a $x^0$-constant 3D hypersurface. 
The arguments of $\varepsilon$ are supposed being given by the Cartesian coordinates.  
It can be seen  that the large U(1) current can be defined as
$J_{\varepsilon}^\mu \equiv \varepsilon \, J_{\rm U(1)}^\mu$.

\vspace{-1.0mm}
%=============================================
\section{The large chiral symmetry}
\label{rymntr}
%=============================================

In the previous section, 
the large U(1) charges were presented on the Cartesian coordinates.
In this section, we heuristically construct the chiral charges based on these. 
We refer to these as \textit{the large chiral charges}.
Then, we show that the transformations generated by these charges are symmetric 
in our model and that they can be defined using Noether's theorem.
\vspace{-1.0mm}

%=============================================
\subsection{Heuristic constitution of the large chiral charges}
\label{rdryq}
%=============================================

The chiral charge can be obtained from the U(1) charge
by changing its integrand $\bar{\psi} \gamma^\mu \psi$ (the U(1) current) to $\psi \gamma^5\gamma^\mu \psi$. 
This means, giving the U(1) charge in the chiral representation
(the form in which the left- and right-hand components are separately presented),
to flip the sign of the right-hand component.
In light of this, we change $Q_{\varepsilon}$ in (\ref{dtre}) to a chiral charge. 

For this purpose, taking the chiral representation as
$\gamma^\mu=
\left(
\begin{array}{cc}
0 & \sigma^\mu \\
\bar{\sigma}^\mu & 0 \\
\end{array}
\right)$ and $\psi=(\xi,\eta)^T$, where 
%--- 
$\sigma^\mu \equiv (I_2, \vec{\sigma})$, 
%---
$\bar{\sigma}^\mu \equiv (I_2, -\vec{\sigma})$ ($\vec{\sigma}$ mean Pauli matrices), 
we can give $Q_{\varepsilon}$ in (\ref{dtre}) as 
\vspace{-1.0mm}
\begin{eqnarray}\label{nlbyj}
\textrm{$Q_{\varepsilon}$ in (\ref{dtre})}
\!\!\!
&=& \hspace{1.4mm}
\overbrace{
\frac{1}{2}
\int_\Sigma d^3x \, \varepsilon\,\Big(
  h_z{}^\mu \,\nabla_\mu         (h_{\bar{z}}{}^\mu F_{\mu t})
+ h_{\bar{z}}{}^\mu \,\nabla_\mu (h_{z}{}^\mu F_{\mu t})
\Big)
- e
\int_\Sigma d^3x \, \varepsilon\, \eta^\dagger \,\eta 
}^{\textrm{Contribution of left-hand (L.H.)}}
\nonumber \\*[0.0mm]
&&\!\!\!\!\!
+ \,
\overbrace{
\frac{1}{2}
\int_\Sigma d^3x \,\varepsilon \,\Big(
  h_z{}^\mu \,\nabla_\mu         (h_{\bar{z}}{}^\mu F_{\mu t})
+ h_{\bar{z}}{}^\mu \,\nabla_\mu (h_{z}{}^\mu F_{\mu t})
\Big) 
- e
\int_\Sigma d^3x \, \varepsilon\, \xi^\dagger \,\xi 
}^{\textrm{Contribution of right-hand (R.H.)}},
\end{eqnarray}
where L.H. and R.H. mean the left- and the right-hand parts, respectively.
The first and second terms in the L.H. and R.H. are usually referred to as "soft-term" and "hard-term".

We perform the manipulation mentioned in the beginning of this subsection to (\ref{nlbyj}), 
which is to change the R.H. as $+\textrm{R.H.} \to -\textrm{R.H.}$, leaving L.H. as it is. 
As a result of this, (\ref{nlbyj}) is changed as follows:
\vspace{-3.0mm}
\begin{eqnarray}\label{gxsrj}
\textrm{L.H.}-\textrm{R.H.}
=
-e\int_\Sigma  d^3x \, \varepsilon\, (\eta^\dagger \,\eta-\xi^\dagger  \,\xi)
= 
\int_\Sigma  d^3x \,  J^\mu_{c\varepsilon}
\equiv Q_{{\rm c}\varepsilon} ,
\quad
J^\mu_{c}            \equiv -e \bar{\psi}\gamma_5\gamma^\mu \psi, \quad
J^\mu_{c\varepsilon} \equiv \varepsilon J^\mu_{c},
\end{eqnarray} 
where we have defined the obtained quantity as $Q_{{\rm c}\varepsilon}$, 
We refer to this as \textit{the large chiral charge}. 
Remarkably, there is no contribution from the soft-term as these cancel each other out.

\vspace{-1.0mm}
%=============================================
\subsection{The large chiral transformations and their charges in terms of Noether's theorem}
\label{irdvs}
%============================================= 

In Sec.\,\ref{rdryq}, the large chiral charges were heuristically constructed. 
In this subsection, we firstly define the transformations generated by them as \textit{the large chiral symmetry}, 
then, obtain the large chiral charges heuristically constructed in Sec.\,\ref{rdryq} 
as the conserved charges based on Noether's theorem. 
\newline

The simultaneous Poisson brackets
for our Dirac field $\psi^\alpha$ (\textit{the Grassmann bracket}) are given as
\vspace{-1.0mm}
\begin{subequations}\label{brwa}
\begin{align} 
\label{brwa1}
\{\psi_\alpha (t,\vec{x}),\pi^\beta (t,\vec{y})\}_{{\rm G.B.}}
&= \delta_\alpha{}^\beta\,\delta^3(\vec{x}-\vec{y}), 
\\*[1.0mm]
\label{brwa2}
\{\psi_\alpha (t,\vec{x}),\psi_\beta (t,\vec{y})\}_{{\rm G.B.}}
&= \{\pi^\alpha (t,\vec{x}),\pi^\beta (t,\vec{y})\}_{{\rm G.B.}} = 0,
\end{align} 
\end{subequations}
where $(t,\vec{x})$ and $(t,\vec{y})$ are the Cartesian coordinates.  
$\{\cdot,\cdot\}_{{\rm G.B.}}$ and $\pi_\alpha$ 
are defined as
\vspace{-1.0mm}
\begin{subequations}\label{svse0u}
\begin{align} 
\label{svse0u1}
\{X(t,\vec{x}),Y(t,\vec{y})\}_{{\rm G.B.}}
\, \equiv& \,
\int d^3x
\left(
\frac{\partial X(t,\vec{x})}{\partial \psi^\alpha (t,\vec{z})} 
\frac{\partial Y(t,\vec{y})}{\partial \pi_\alpha(t,\vec{z})}
+ 
\frac{\partial X(t,\vec{x})}{\partial \pi_\alpha(t,\vec{z})} 
\frac{\partial Y(t,\vec{x})}{\partial \psi^\alpha(t,\vec{z})}\right), \\*[1.0mm]
\label{svse0u2}
\pi^\alpha (t,\vec{x}) 
\, \equiv& \,\,
\frac{\partial \,{\cal L}_0}{\partial \,(\partial_0 \psi_\alpha (t,\vec{x}))}
=i \psi^\dagger{}^\alpha (t,\vec{x}).
\end{align} 
\end{subequations}
$X$ and $Y$ are some functions and ${\cal L}_0$ is given in (\ref{erer}). 
With (\ref{svse0u2}), the following Grassmann bracket can be obtained from (\ref{brwa1}): 
\vspace{-1.0mm}
\begin{eqnarray} \label{sndk}
\{\psi_\alpha (t,\vec{x}),\psi^\dagger{}^\beta (t,\vec{y})\}_{{\rm G.B.}}
= -i\delta_\alpha{}^\beta\,\delta^3(\vec{x}-\vec{y}).
\end{eqnarray} 
Using (\ref{brwa2}), (\ref{sndk}) 
and $Q_{{\rm c},\varepsilon}$  in (\ref{gxsrj}),
the following Grassmann bracket can be obtained: 
\vspace{-1.0mm}
\begin{align}\label{w5srb}
\{Q_{{\rm c}\varepsilon}  ,\psi\}_{{\rm G.B.}}       = i\gamma_5 \,\varepsilon \, \psi, \quad 
\{Q_{{\rm c}\varepsilon}  ,\bar{\psi}\}_{{\rm G.B.}} = i\bar{\psi}\gamma_5 \,\varepsilon.
\end{align}  
(\ref{w5srb}) means that, 
if $Q_{{\rm c}\varepsilon}$ are conserved quantities, 
$Q_{{\rm c}\varepsilon}$ are the charges associated with the following  transformations: 
\vspace{-2.0mm}
\begin{subequations}\label{zery}
\begin{align} 
\label{zery1}
\psi       \to & \,\,\psi'       = e^{ie\varepsilon\lambda\,\gamma_5} \,\psi 
                                 = \psi      +ie \, \varepsilon\lambda \, \gamma_5\psi        
                                             +{\cal O}(\lambda^2), 
\\*[1.0mm]
\label{zery2}
\bar{\psi} \to & \,\,\bar{\psi}' = \bar{\psi}\,e^{ie\varepsilon\lambda\,\gamma_5}
                                 = \bar{\psi}+ie\,  \varepsilon\lambda \, \bar{\psi}\gamma_5  
                                             +{\cal O}(\lambda^2).
\end{align} 
\end{subequations}
We refer to these as \textit{the large chiral transformations}.
\newline

We verify that (\ref{zery}) is symmetric transformation in our model (\ref{erer}). 
For this purpose, we examine the variation arising in (\ref{erer}) by (\ref{zery}):
\vspace{-1.0mm}
\begin{eqnarray}\label{eerba}
\int d^4x \,\bar{\psi} i\!\slash{D}\psi \to
\int d^4x \,\bar{\psi}'\,i\slash{D}\,\psi'
\!\!\! &=& \!\!\!
   \int d^4x \, \bar{\psi\,}i\slash{D}\psi
- e\int d^4x \, \partial_\mu(\lambda\,\varepsilon)\, \bar{\psi}\gamma_5\gamma^\mu \psi
\nonumber\\*[1.0mm]
\!\!\! &=& \!\!\!
  \int d^4x \, \bar{\psi\,}i\slash{D}\psi
+ \underbrace{
\int d^4x \, \partial_\mu (\lambda \, \varepsilon J^\mu_c)
}_{=0}
- \underbrace{
\int d^4x \, \lambda\,\varepsilon\,\partial_\mu J^\mu_c
}_{=0},
\end{eqnarray}
where $J^\mu_{c}$ is defined in (\ref{gxsrj}). 
Since $\lambda$ is assumed to vanish at $r \to \infty$,
the second term vanishes.
The third term also vanishes as 
$J^\mu_{c}$ is constant
due to the chiral symmetry. 
Therefore, (\ref{zery}) is a symmetric transformation in our model (\ref{erer}). 
\newline

Now, we show that $Q_{{\rm c}\varepsilon}$ are the charges 
associated with the large chiral transformations (\ref{zery}) 
based on Noether's theorem.  
For (\ref{zery}), the quantities in (\ref{gdwr}) can be written as follows:
\vspace{-2.0mm}
\begin{subequations}\label{eaeha}
\begin{align} 
\label{eaeha1}
X^\mu  = & \, N^\mu{}_{,A}  = \textrm{$0$ for all $\mu$ and $A$}
\\*[1.0mm]
\label{eaeha2}
(M_{,0},M_{,1},M_{,2},M_{,3},M_{,4},M_{,5})=& \, 
(0,0,0,0,ie\varepsilon\gamma_5\psi,ie\varepsilon\bar{\psi}\gamma_5),
\end{align} 
\end{subequations}
where $(\phi_0,\phi_1,\phi_2,\phi_3,\phi_4,\phi_5)=(A_0,A_1,A_2,A_3,\psi,\bar{\psi})$ 
and $n$ is $1$, which is $r_{\rm max}$; 
therefore we omitted to denote $r$.
Using these in the same way as (\ref{j5ir}), $B^\mu$ and $C^{\mu\nu}$ in (\ref{resyu}) are obtained as 
\vspace{-1.0mm}
\begin{eqnarray}\label{jswzr}
B^\mu = J^\mu_{c\varepsilon}, \quad 
C^{\mu\nu} = 0.
\end{eqnarray}
Therefore, the following conserved law can be obtained by following (\ref{resyu}):
\vspace{-1.0mm}
\begin{eqnarray}\label{yofa}
\partial_\mu J^\mu_{c\varepsilon} =0.
\end{eqnarray}
From this, $Q_{{\rm c}\varepsilon}$ in (\ref{gxsrj}) can be obtained 
as the charges for the large chiral transformations (\ref{zery}). 

\vspace{-1.0mm}
%=============================================
\subsection{The large U(1) gauge transformations}
\label{irvrn}
%============================================= 

In the previous subsection, the large chiral transformations of the fermionic and gauge fields are given. 
In this subsection, the large U(1)  transformations of the fermionic field are given.
From (\ref{w5srb}), 
\vspace{-1.0mm}
\begin{align}\label{wryb}
\{Q_{\varepsilon}  ,\psi\}_{{\rm G.B.}}       = i\,\varepsilon \, \psi, \quad 
\{Q_{\varepsilon}  ,\bar{\psi}\}_{{\rm G.B.}} = i\,\varepsilon \, \bar{\psi}.
\end{align}  
From these, the large transformed fermionic field and gauge field can be known as follows:
\vspace{-1.0mm}
\begin{align}\label{ukuy}
\psi       \to \,\psi'       = e^{+ie\varepsilon\lambda} \,\psi, \quad
\bar{\psi} \to \,\bar{\psi}' = e^{-ie\varepsilon\lambda} \,\bar{\psi}, \quad
A_\mu      \to \,\, A_\mu+ e^{-1}\,\partial_\mu (\varepsilon\lambda).
\end{align} 
In the one above, 
the transformation of $A_\mu$ cannot be obtained from the analysis with the Poisson bracket (P.B.). 
Thus, we have obtained it by deducing from the transformations of $\psi$ and $\bar{\psi}$.
This is a general situation 
by the reason noted in the footnote\footnote{
%========================
%======= FOOTNOTE =======
%========================
In the global transformation, 
charges generate transformations as 
$\{Q_r^{\rm (G)},\phi_A\}_{\rm P.B.}= M_{r,A}$ 
($M_{r,A}$ and $Q_r^{\rm (G)}$ are those in (\ref{redz}) and (\ref{trshk}), respectively).
%---
Then, as can be seen from (\ref{trshk}) and (\ref{rrwk}), 
the forms of the charges in the global and local transformation are the same as each other. 
From this, it can be seen that the charges in the local transformation can generate 
the transformations that the charges in the global transformation  generate.  
This means $\{Q_r^{\rm (L)},\phi_A\}_{\rm P.B.}= M_{r,A}$
(this $M_{r,A}$  and $Q_r^{\rm (L)}$ are those in (\ref{gdwr}) and (\ref{rrwk}), respectively).  

Concretely, the transformations generated by $\{Q_r^{\rm (L)},\phi_A\}_{\rm P.B.}$ 
are the phases of the matter fields (such as (\ref{vwrer2})),
and the transformations on the gauge fields (such as (\ref{vwrer1})) are not generated
by the charges in the gauge transformations. 
}.
%========================
%========================
%========================

\vspace{-1.0mm}
%=============================================
\section{The anomaly equation for the large chiral symmetry}
\label{ykiwf}
%=============================================  

In the previous section, 
we obtained the conservation law (\ref{yofa})
associated with the large chiral symmetry (\ref{zery}). 
In general, conservation laws including $\gamma_5$ get quantum corrections, i.e. anomalies.
Therefore, in this section, we provide (\ref{yofa}) including anomalies.
\newline

We can give (\ref{yofa}) as follows:
\vspace{-1.0mm}
\begin{eqnarray}\label{ytka}
\partial_\mu \langle J^\mu_{c\varepsilon} \rangle 
=
(\partial_\mu \varepsilon) \, \langle J^\mu_{c} \rangle
+\varepsilon \, \partial_\mu \langle J^\mu_{c} \rangle
=
(\partial_\mu \varepsilon) \, \langle J^\mu_{c} \rangle
-e\,\varepsilon \, \partial_\mu \langle \bar{\psi} \gamma_5\gamma^\mu \psi \rangle,
\end{eqnarray}
where $J^\mu_{c\varepsilon}$ and $J^\mu_{c}$ are defined in (\ref{gxsrj}).
$\langle \cdots \rangle$ means the quantum e.v. defined as  
\vspace{-1.0mm}
\begin{eqnarray} \label{dalb}
\label{dalb1}
\langle {\cal O} \rangle
\equiv 
\int 
{\cal D}\psi {\cal D} \bar{\psi} \,
{\cal D}A_\mu {\cal D} B \,
{\cal D}c {\cal D} \bar{c} \,\,\, {\cal O}\, 
\exp  \, [i\int d^4x \sqrt{-g} \, ({\cal L}_0 +{\cal L}_{\rm g.f.})]. 
\end{eqnarray} 
${\cal L}_{\rm g.f.}$ means some gauge-fixing terms.
While the first term in (\ref{ytka}) cannot be generally calculated more than that, 
the result of the second term is known, using which, (\ref{ytka}) can be given as follows:
\vspace{-1.0mm}
\begin{eqnarray}\label{reois}
\partial_\mu \langle J^\mu_{c\varepsilon} \rangle
= (\partial_\mu \varepsilon) \, \langle J^\mu_c \rangle
-\varepsilon \, i \,\frac{\hbar \, e^3}{16\pi^2}\,\varepsilon^{\mu \nu \rho \sigma}F_{\mu \nu} F_{\rho \sigma}. 
\end{eqnarray}
In this study, $\hbar$ is taken to 1.
This is the anomaly equation with regard to the large chiral symmetry. 

\vspace{-1.0mm}
%=============================================
\section{The anomaly equation from the large BRS-transformed effective action}
\label{ypvsb}
%=============================================  

In the previous section, 
we obtained the anomaly equation
with regard to the large chiral symmetry 
as can be seen in (\ref{reois}).
%---
In this section, 
we derive it from the evaluation of the one-loop diagrams.
 
Specifically, 
we first define the BRS transformations of the large U(1) symmetry (\textit{the large BRS transformations}) (Sec.\,\ref{ypvsb2}). 
%---
Then, by performing these large BRS transformations 
to the one-loop diagrams constituting the effective action of our model (\ref{erer}),
we once obtain anomaly equations associated with the large U(1) current (Sec.\,\ref{ydtyj2}).
%--- 
Then, we change the anomaly equation obtained in Sec.\,\ref{ydtyj2} 
by axializing these one-loop diagrams 
(by changing one of the vector type vertices in each of these one-loop diagrams to the axial-vector type) (Sec.\,\ref{ypvsb3}).
%---
Then, evaluating each of these axialized one-loop diagrams, 
we derive the anomaly equation in (\ref{reois}) (Sec.\,\ref{hbrbes}\,-\,\ref{hbddT5}). 

%=============================================
\subsection{The BRS transformation of the large U(1) gauge transformation}
\label{ypvsb2}
%=============================================  

Following the general procedure to define a BRS transformation for a given gauge transformation\footnote{
%====================
%===== FOOTNOTE =====
%====================
It is just a replacement of the gauge transformation parameter
with a product of a Grassmann number and a ghost field such as
$\lambda \to \theta c$ 
($\theta$ means a Grassmann number),
then take up the part other than $\theta$.
%====================
%====================
%====================
}, 
we define the BRS transformation of the large gauge transformation 
from (\ref{ukuy}) as 
\vspace{-1.0mm}
\begin{eqnarray}\label{fytou}
\boldsymbol{\delta}_{\rm \varepsilon} \psi(x)  = i e\,\varepsilon(x) c(x) \psi(x), \quad
\boldsymbol{\delta}_{\rm \varepsilon} A_\mu(x) = e^{-1}\partial_\mu  (\varepsilon(x) c(x)),
\end{eqnarray}
where $c(x)$ is the ghost field.
The coordinates are Cartesian.
We refer to these as \textit{the large BRS transformations} and denote them with 
``$\boldsymbol{\delta}_{\rm \varepsilon}$''. 
These are given in the momentum space as 
\vspace{-1.0mm}
\begin{eqnarray}\label{vbpv}
\boldsymbol{\delta}_{\rm \varepsilon} \psi(k)  = i e\,\varepsilon (k) c (k) \psi (k), \quad
\boldsymbol{\delta}_{\rm \varepsilon} A_\mu(k) = -ie^{-1}k_\mu \varepsilon (k) c(k),
\end{eqnarray}
where 
$\varepsilon(x) c(x) \psi(x)         = \int \!\frac{d^4x}{(2\pi)^4}\,e^{-ikx}\, \varepsilon(k) c(k) \psi(k)$ and 
$\varepsilon(x) c(x)                 = \int \!\frac{d^4x}{(2\pi)^4}\,e^{-ikx}\, \varepsilon(k) c(k)$.

\vspace{-1.0mm}
%=============================================
\subsection{The anomaly equation of the large U(1) current} 
\label{ydtyj2}
%=============================================  

To derive the anomaly equation of the large chiral current,
we first derive the anomaly equation of the large U(1) current
from the breaking of the large BRS symmetry (\ref{fytou}) in the effective action of (\ref{erer}).
We begin with the definition of the effective action:
\vspace{-1.0mm}
\begin{eqnarray}\label{tatky}
e^{i\Gamma[A]}
=\int \! {\cal D}\psi{\cal D}\bar{\psi}\exp \,[i\int \! d^4x  \sqrt{-g}\, {\cal L}_0].
\end{eqnarray}
${\cal L}_0$ is given in (\ref{erer}).   
Since the spacetime is flat and the coordinates are Cartesian, $\sqrt{-g}=1$. 
This effective action is with regard to $A_\mu$ and can be presented as
\vspace{-1.0mm}
\begin{eqnarray}\label{rtoiu}
\Gamma[A] 
= 
i^{-1} \, \Tr \,\ln [{\slash \,\partial}-ie{\slash A}]
- \int \! d^4x \,\frac{1}{4}\,F_{\mu\nu}F^{\mu\nu},
\end{eqnarray} 
where 
$
\int \! {\cal D}\psi{\cal D}\bar{\psi}\exp \,[i\int \! d^4x  \, \bar{\psi}i{\slash D}\psi]
=\Det(-\gamma^0{\slash D})$.  
$\Tr$ and $\Det$ mean the trace and determinant for the coordinates and the indices of fields. 

Operating $\boldsymbol{\delta}_{\rm \varepsilon}$ to 
the effective action ($\boldsymbol{\delta}_{\rm \varepsilon}$ is given in (\ref{fytou})),
\vspace{-1.0mm}
\begin{eqnarray}\label{wqrrl0}
\boldsymbol{\delta}_{\rm \varepsilon} \Gamma[A]
\!\!\! &=& \!\!\! 
\int \! d^4x \, \boldsymbol{\delta}_{\rm \varepsilon} A_\mu \, 
\frac{\delta \Gamma[A]}{\delta A_\mu} 
=
\int \! d^4x \, \boldsymbol{\delta}_{\rm \varepsilon} A_\mu \, 
\frac{\delta }{\delta A_\mu}
\Big( 
i^{-1} \,\ln \int \! {\cal D}\psi{\cal D}\bar{\psi}\exp \,[i\int \! d^4x \, {\cal L}_0]
\Big)
\nonumber\\*[1.0mm]
\!\!\! &=& \!\!\!
i^{-1}\int \! d^4x\, \boldsymbol{\delta}_{\rm \varepsilon} A_\mu \,  
\Big( 
\! -i\langle J_{\rm U(1)}^\mu \rangle 
+i\,\langle \partial_\nu  F^{\nu \mu} \rangle 
\Big)
=
-\int \! d^4x\, \boldsymbol{\delta}_{\rm \varepsilon} A_\mu \,  
\langle J_{\rm U(1)}^\mu \rangle,
\end{eqnarray}
where $\partial_\nu F^{\nu \mu}=0$ by the equations of motion.
With (\ref{fytou}), 
\vspace{-1.0mm}
\begin{eqnarray}\label{wqrrl}
\textrm{r.h.s. of (\ref{wqrrl0})}
=
-e^{-1}\int \! d^4x \, (
  \partial_\mu \varepsilon \, c 
+ \varepsilon \,  \partial_\mu c 
)\, \langle J_{\rm U(1)}^\mu \rangle  
=
-e^{-1}
\int \! d^4x \, c \, 
\Big( \, 
\!   \partial_\mu \varepsilon \,  \langle J_{\rm U(1)}^\mu \rangle 
   - \partial_\mu (\varepsilon \,  \langle J_{\rm U(1)}^\mu \rangle )
\Big),
\end{eqnarray}
where supposing $c$ vanishes at $r \to \infty$,  
$\int d^4x \,  \partial_\mu(\varepsilon c \langle J_{\rm U(1)}^\mu \rangle  )=0$.  
Therefore, (\ref{wqrrl0}) is given as follows:
\vspace{-1.0mm}
\begin{eqnarray}\label{reun}
\boldsymbol{\delta}_{\rm \varepsilon} \Gamma[A]
=-e^{-1}
\int \! d^4x \, c \, 
\Big(
   \partial_\mu \varepsilon \,  \langle J_{\rm U(1)}^\mu \rangle 
-  \partial_\mu \langle J_{\varepsilon}^\mu \rangle \Big),
\end{eqnarray}
where $\varepsilon \,   J_{\rm U(1)}^\mu \equiv J_{\varepsilon}^\mu$ 
as defined in (\ref{dtre}).
From this, we can obtain the equation 
in the form $\partial_\mu \langle J_{\varepsilon}^\mu \rangle= \cdots$,
which is the anomaly equation with regard to  the large U(1) gauge symmetry.  

\vspace{-1.0mm}
%=============================================
\subsection{Change the anomaly equation (\ref{reun}) 
to the one with regard to the large chiral current}
\label{ypvsb3}
%=============================================  

\begin{figure}[H]  
\vspace{-10.5mm} 
\begin{center}
\includegraphics[clip,width=10.6cm,angle=-90]{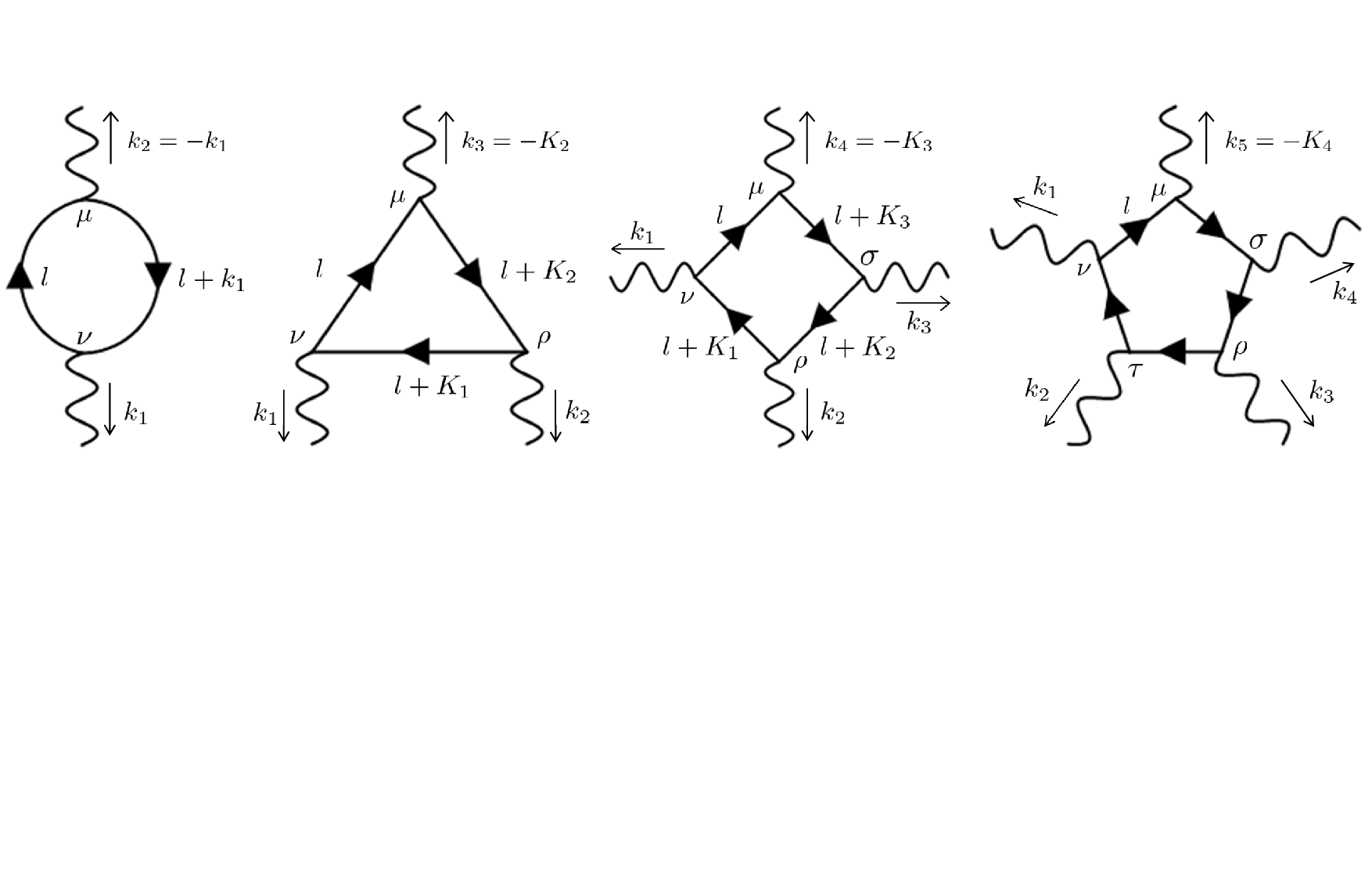} 
\end{center}
\vspace{-53.0mm}
\caption{
The Feynman diagrams of $\Gamma^{(n)}$ for $n=2,3,4,5$ in (\ref{yvdbe2}).}
\label{wwewo}
\end{figure}

As such, we change the equation for $\partial_\mu \langle J_{\varepsilon}^\mu \rangle$ 
to the one with regard to the large chiral current in (\ref{gxsrj}).
For this purpose, we first  give the effective action (\ref{rtoiu}) 
to the one-loop order by expanding it:
\vspace{-1.0mm}
\begin{eqnarray}\label{rau0}
i\,\Gamma[A] 
= 
\Tr \, \Big[ 
\ln [{\slash \,\partial}]
-\sum_{n=1}^\infty \frac{1}{n} \Big({\slash \,\partial}^{-1} \, ie{\slash A} \Big)^n 
\Big]
%-----
= 
\Tr \, \ln [{\slash \,\partial}]
+\sum_{n=1}^\infty \Gamma[A] \big\vert_{\textrm{$n$-th}}
+\cdots, 
\end{eqnarray}
\vspace{-5mm}
\begin{subequations}\label{yvdbe}
\begin{align}
\label{yvdbe1}
\textrm{where} \qquad
\Gamma[A] \big\vert_{\textrm{$n$-th}}
\equiv& 
\int \Big(\prod_{i=1}^n\frac{d^4k_i}{(2\pi)^4}\Big)
\, (2\pi)^4 \, \delta^4 (K_n)
\, \frac{1}{n} \, \Big(\prod_{i=1}^n eA_{\mu_n}(-k_n) \Big) \, \Gamma^{(n)}, 
\\* 
%==========
\label{yvdbe2}
\Gamma^{(n)} 
\equiv& \,\,
\Gamma^{\mu_1 \cdots \mu_n} 
\equiv
-\int \frac{d^4l}{i(2\pi)^4}
\, \tr \, 
\Big[
\frac{1}{-{\slash ~ l}} \, \gamma^{\mu_n}
\prod_{i=n-1}^1
\frac{1}{-({\slash ~ l}+{\slash ~ K_i})} \, \gamma^{\mu_{i}} 
\Big].
\end{align}
\end{subequations}
``$\cdots$'' means the contribution from more than two-loop order, and
$K_n \equiv \sum_{i=1}^n k_i$. 
$\Gamma^{(n)}$ is just an abbreviation of $\Gamma^{\mu_1 \cdots \mu_n}$. 
$\tr$ means the trace for spinor's indices. 
In giving $\Tr$ by an integral in (\ref{yvdbe}), 
we used the Feynman rules\footnote{
%==============================
%========== FOOTNOTE ==========
%==============================
For the model (\ref{erer}), 
{\bf (i)} $(m-{\slash k}-i\epsilon)^{-1}$ corresponds to the fermionic inner-lines with momentum $k_\mu$ ($m=\epsilon=0$),
{\bf (ii)} $e \gamma^\mu A_\mu$ corresponds to the vertexes, and
{\bf (iii)} $-\int \frac{d^4l}{i(2\pi)^4}$ corresponds to the loop-integral the fermions run around. 
%==============================
%==============================
%==============================
}.
We ignored $- \int d^4x \,\frac{1}{4}\,F_{\mu\nu}F^{\mu\nu}$ in (\ref{rtoiu})
as it can be ignored as seen under (\ref{wqrrl0}).
We show the Feynman diagrams of (\ref{yvdbe2}) for $n=2,3,4,5$ in Fig.\,\ref{wwewo}.
%==================
%===== FIGURE =====
%==================
%==================
%==================
%==================

We have derived an anomaly equation for the large U(1) current in (\ref{reun}). 
However, our aim is the anomaly equation associated with the large chiral current. 
%---
In addition, even if we could eventually obtain
$\partial_\mu \langle J^\mu \rangle$ and 
$\partial_\mu \langle J^{5\mu} \rangle$
from (\ref{reun}), they would not still immediately coincide with the right ones.
%---
We explain this point by referring to the already known U(1) case.

Suppose the effective action (\ref{tatky}) to be separately given by the chirality as 
$e^{i\Gamma[A]}=\Det (i {\slash D}_{\rm R})\,\Det (i {\slash D}_{\rm L})$. 
Then, from the equation analogous  to (\ref{reun}),
$\partial_\mu \langle J_{{\rm L}}^\mu \rangle$ and 
$\partial_\mu\langle J_{{\rm R}}^\mu \rangle$ 
can be obtained. 
%---
From these, 
$\partial_\mu \langle J^\mu \rangle$ and 
$\partial_\mu \langle J^{5\mu} \rangle$
can be obtained as 
$\partial_\mu \langle J_{{\rm L}}^\mu \rangle+\partial_\mu \langle J_{{\rm R}}^\mu \rangle$ and
$\partial_\mu \langle J_{{\rm L}}^\mu \rangle-\partial_\mu \langle J_{{\rm R}}^\mu \rangle$, respectively.  
%---
However, this $\partial_\mu\langle J^\mu \rangle$ is non-zero
and $\partial_\mu\langle J^{5\mu} \rangle$ does not coincide with the right chiral anomaly.  

However, there is degree of freedom in the effective action to modify it
keeping the correspondence of the effective action with the original theory as it is (as the matter of how to take the regularization term), 
and utilizing that, the effective action can be modified  
in such a way that $\partial_\mu\langle J^\mu \rangle=0$ can be realized. 
%---
At this time, $\partial_\mu\langle J^{5\mu} \rangle$ coincides with the right chiral anomaly. 
However, the analysis of this modification is highly dense\footnote{
%====================
%===== FOOTNOTE =====
%====================
As the preparation to perform that modification, 
we first need to analyze Wess-Zumino consistency-condition. 
%---
Also, we need to prepare various mathematical techniques 
(the superfield formalism, Cartan homotopy formula, Chern-Simon forms, etc)
to smoothly perform the analysis. 
%---
We will take care of these in the future work. 
%====================
%====================
%====================
}. 

Since this situation would be the same 
in the case of obtaining the anomaly equation of the large chiral current, 
and as long as we begin with the effective action (\ref{tatky}),
we would have to perform the modification mentioned above. 
However, since it is beyond the issue this study addresses, 
we will obtain $\partial_\mu \langle J^{5\mu} \rangle$ in another way. 

That another way is to axialize each of the one-loop diagrams (\ref{yvdbe2})
by changing a $\gamma^\mu$ in the each by hands as follows:
\vspace{-5.0mm}
\begin{eqnarray}\label{iukne}
\gamma^\mu \to \gamma^\mu \gamma_5.
\end{eqnarray} 
By this, $J_{\rm U(1)}^\mu$ and $J_{\varepsilon}^\mu$ in the r.h.s. of (\ref{reun}) 
are changed to $J_{\rm c}^\mu$ and $J_{c\varepsilon}^\mu$
($J_{\rm U(1)}^\mu \equiv -e\bar{\psi}         \gamma^\mu \psi$ and 
 $J_{\rm c}^\mu    \equiv -e\bar{\psi} \gamma_5\gamma^\mu \psi$, 
and 
$J_{\rm \varepsilon}^\mu  \equiv \varepsilon J_{\rm U(1)}^\mu$ and
$J_{\rm c\varepsilon}^\mu \equiv \varepsilon J_{\rm c}^\mu$
as can be seen under (\ref{dtre}) and in (\ref{gxsrj})). 
\noindent
Therefore, performing (\ref{iukne}) to (\ref{reun}), 
we can obtain the equation to give $\partial_\mu  \langle J_{c\varepsilon}^\mu \rangle$ 
by a calculable quantity $\boldsymbol{\delta}_{\rm \varepsilon} \Gamma_{\cal A}[A]$ as follows:
\vspace{-1.0mm}
\begin{eqnarray}\label{efgnds}
\boldsymbol{\delta}_{\rm \varepsilon} \Gamma_{\cal A}[A]
=
-e^{-1} \int d^4x \, c \, 
\Big(
  \partial_\mu \varepsilon \,  \langle J_{\rm c}^\mu \rangle 
- \partial_\mu  \langle J_{c\varepsilon}^\mu \rangle 
\Big).
\end{eqnarray}
\vspace{-7mm}
\begin{eqnarray}\label{sytst}
\textrm{where} \qquad
 i \,\Gamma_{\cal A}[A] 
\!\!\! &\equiv& \!\!\!
\Tr \, \ln \,[{\slash \,\partial}]
+\sum_{n=1}^\infty \Gamma_{\cal A}[A] \big\vert_{\textrm{$n$-th}}, 
\\* 
%-----
\label{yjspe1}
\Gamma_{\cal A}[A] \big\vert_{\textrm{$n$-th}}
\!\!\! &\equiv& \!\!\!
\int \Big(\prod_{i=1}^n\frac{d^4k_i}{(2\pi)^4}\Big)
\, (2\pi)^4 \, \delta^4 (K_n )
\, \frac{1}{n} \, \Big(\prod_{i=1}^n eA_{\mu_i}(-k_i) \Big) \, \Gamma_{\cal A}^{(n)}, 
\nonumber\\*  
%-----
\label{yjspe2}
\Gamma_{\cal A}^{(n)} 
\!\!\! &\equiv& \!\!\!
\Gamma_{\cal A}^{\mu_1 \cdots \mu_n} 
\equiv
-\int \frac{d^4l}{i(2\pi)^4} \, \tr \,
\Big[
\frac{1}{-{\slash ~ l}} \, \gamma^{\mu_n} \gamma_5 \,
\prod_{i=n-1}^1
\frac{1}{-({\slash ~ l}+{\slash ~ K_i})} \, \gamma^{\mu_i} 
\Big].
\nonumber
\end{eqnarray}
The equation $\partial_\mu  \langle J_{c\varepsilon}^\mu \rangle =\cdots$
to be obtained from this is what we want to obtain, and the final result is given in (\ref{srds}).
With (\ref{sytst}), the l.h.s. of (\ref{efgnds}) can be written as
\vspace{-1.0mm}
\begin{subequations}\label{hbdd}
\begin{align}
\boldsymbol{\delta}_{\rm \varepsilon} \, \Gamma_{\cal A}[A] \vert_{\textrm{$n$-th}}
=& \,
\sum_{n=1}^\infty\int \Big(\prod_{i=1}^n\frac{d^4k_i}{(2\pi)^4}\Big)
\, (2\pi)^4 \, \delta^4 (K_n)
\, \Big(e\,\boldsymbol{\delta}_{\rm \varepsilon}(A_{\mu_n}(-k_n)) 
\prod_{i=1}^{n-1} 
eA_{\mu_i}(-k_i) \Big) \, \Gamma_{\cal A}^{(n)} 
\nonumber\\*
%==========
\label{hbdd1}
= 
& \hspace{4mm}
\int \frac{d^4k}{(2\pi)^4}
\,(2\pi)^4 \, \delta^4(k_1)\, 
\Big( \varepsilon(-k_1)\, c(-k_1)\Big) \, ik_{1\mu} \, \Gamma^{(1)}_{\cal A} 
\\*
%==========
\label{hbdd2}
& +
\int \Big( \prod_{i=1}^2\frac{d^4k_i}{(2\pi)^4} \Big)
(2\pi)^4 \, \delta^4 ( K_2 ) \,
\Big( \varepsilon(-k_2)\, c(-k_2) \, eA_{\mu_1}(-k_1) \Big) 
\, ik_{2\mu} \, \Gamma^{(2)}_{\cal A} 
\\*
%==========
\label{hbdd3}
& +
\int \Big( \prod_{i=1}^3\frac{d^4k_i}{(2\pi)^4} \Big)
(2\pi)^4 \, \delta^4 ( K_3 ) \,
\Big( \varepsilon(-k_3)\, c(-k_3)\, \prod_{i=1}^2 eA_{\mu_{n}}(-k_{n})\Big) 
\,ik_{3\mu} \,\Gamma^{(3)}_{\cal A} 
\\*
%==========
\label{hbdd4}
& +
\int \Big( \prod_{i=1}^4 \frac{d^4k_i}{(2\pi)^4} \Big)
(2\pi)^4 \, \delta^4 ( K_4 ) \,
\Big( \varepsilon(-k_4)\, c(-k_4) \, \prod_{i=1}^3 eA_{\mu_{n}}(-k_{n})\Big) 
\,ik_{4\mu} \,\Gamma^{(4)}_{\cal A} 
\\*
%==========
\label{hbdd5}
& +
\int \Big( \prod_{i=1}^5 \frac{d^4k_i}{(2\pi)^4} \Big)
(2\pi)^4 \delta^4 ( K_5 ) \, 
\Big( \varepsilon(-k_5)\, c(-k_5) \, \prod_{i=1}^4 eA_{\mu_{n}}(-k_{n}) \Big)
\,ik_{5\mu} \,\Gamma^{(5)}_{\cal A} 
\\*
%==========
\label{hbdd6}
&+ \cdots.
\end{align}
\end{subequations}
Since $\Gamma^{(n)}$ for $n=1,\cdots,5$ are diverged,
we explicitly wrote to $n=5$
(as can be seen from (\ref{sytst}), the divergence-order of  $\Gamma^{(n)}$ is $4-n$, 
but $\Gamma^{(5)}$ potentially includes  log-divergence by the regularization scheme we employ 
as we comment under (\ref{aavsy})). 
``$\cdots$'' means the terms with $n=6,7,\cdots$ which definitely vanish.  
In the following subsections, we evaluate each term.  
The final result is given in (\ref{srds}).

\vspace{-1mm}
%=============================================
\subsection{The calculation of $ik_{3\mu}\Gamma^{(3)}_{\cal A}$ in (\ref{hbdd3})}
\label{hbrbes}
%=============================================  

We now calculate (\ref{hbdd3}), which is given by an $l$-integral.
Since it is UV-diverged, we employ dimensional regularization.
Thus, $4$ of the four-dimensional integral is analytically continued to a complex number $n$. 
We comment on our treatment for this $n$, below. 
%-------------------------
\vspace{-1.0mm}
\begin{itemize} % --(A)
\item 
$n$ is taken to a complex number when the $l$-integral is performed.  
However, in all other calculations, $n$ is treated as a real number.
Here, we will not perform  Euclideanization. 

\item 
The fermionic field is integrated out 
before the $l$-integral 
by supposing that as the one on the real $n$-dimensional Minkowski spacetime.

\item 
$\gamma$-matrices are supposed to be on the real $n$-dimensional Minkowski spacetime
except for the time that $n$ is taken to a complex number in the $l$-integral. 
When $n$ is taken to a complex number, 
$\gamma$-matrices are treated formally,
and concrete calculations with $\gamma$-matrices, 
such as $\tr(\gamma_5 \gamma^\nu \gamma^\rho \gamma^\alpha \gamma^\beta)=-4i\,\epsilon^{\nu \rho \alpha \beta}$, 
are performed after performing the $l$-integral and the $n \to 4$ limit.

\item 
Even when the real $n$-dimensional spacetime is considered,
$\epsilon^{\mu\nu\sigma\rho}$ and $\gamma_5$ work 
only on the four-dimensional part in the real $n$-dimensional spacetime.
%---
Therefore, for example, the commutation relations between $\gamma^\mu$ and $\gamma_5$ 
on the real $n$-dimension are given as follows:
\vspace{-1.0mm}
\begin{eqnarray}\label{stkpv}
\gamma_5 \gamma^\mu=
\left\{
\begin{array}{ll}
-\gamma_5 \gamma^\mu & \textrm{for $\mu=0,1,2,3$,} \\*[1.0mm]
+\gamma_5 \gamma^\mu & \textrm{for $\mu=$ other than $0,1,2,3$.} 
\end{array}
\right.
\end{eqnarray}
\end{itemize}   % --(A)
%-------------------------

\vspace{-1.0mm}
Then, let us first give the integral variable $l_\mu$ on the $n$-dimension as 
\vspace{-1.0mm}
\begin{eqnarray}\label{dweklv}
l_\mu=\uline{l}\hspace{0.4mm}{}_{\mu}+r_\mu,
\end{eqnarray}
where 
$\uline{l}_{\hspace{0.3mm}\mu} \equiv (l_0,l_1,l_2,l_3,0,\cdots,0)$,
$r_\mu                         \equiv (0,0,0,0,l_4,l_5,\cdots ,l_{n-1})$ 
and $l_\mu                     \equiv (l_0,\cdots,l_{n-1})$.
\begin{itemize} % --(B)
\item
However, even in the $n$-dimension,
%--------------------
\vspace{-1.0mm}
\begin{itemize}
\item
The first four elements of the external line's momenta remain as they are, 
and all other elements are 0;  
namely, $k_\mu=(k_0,k_1,k_2,k_3, 0,\cdots,0)$, where $k_0,\cdots,k_3$ are the original values.
\item
The same applies to the external gauge field $A_\mu$; 
namely, $A_\mu=(A_0,A_1,A_2,A_3, 0,\cdots,0)$, where $A_0,\cdots,A_3$ are the original values.
\end{itemize} 
%--------------------
\end{itemize}  % --(B)

Now, we give $k_{3\mu}$ as the difference 
between the incoming and outgoing momenta for the vertex; namely, $k_{3\mu}=l_\mu-(l+K_2)_\mu$.
Using this in  $-k_{3\mu} \gamma^\mu \gamma_5$, 
\vspace{-1.0mm}
\begin{eqnarray}\label{tksjy}
k_{3\mu} \gamma^\mu \,\gamma_5 
=
{\slash ~ l}  \,\gamma_5 
+ \gamma_5 \,({\slash ~ l}+{\slash ~ K_2})
-2 {\slash ~ r}\, \gamma_5,
\end{eqnarray}
where ${\slash ~ l} \gamma_5=-\gamma_5 {\slash ~ l} +2{\slash ~ r} \gamma_5$ by following (\ref{stkpv}). 
Let us rewrite $ik_\mu \Gamma_{\cal A}^{(3)}$ with this. 
Originally, it is given as
\vspace{-1.0mm}
\begin{eqnarray}\label{ttshs}
ik_3{}_\mu \Gamma_{\cal A}^{(3)} 
= 
-\int \! \frac{d^4l}{i(2 \pi)^4} 
\, \tr \,
\Big[ 
\frac{1}{-{\slash ~ l}}                  \,(ik_3{}_\mu \gamma^\mu \gamma_5) \, 
\frac{1}{-({\slash ~ l}+{\slash ~K_2})}  \gamma^\rho 
\frac{1}{-({\slash ~ l}+{\slash ~ K_1})} \gamma^\nu
\Big],
\end{eqnarray}
and we substitute (\ref{tksjy}) into $ik_3{}_\mu \gamma^\mu \gamma_5$. Consequently,
\vspace{-3.0mm}
\begin{eqnarray}\label{asuis}
ik_3{}_\mu \widetilde{\Gamma}_{\cal A}^{(3)} 
\!\!\! &=& \!\!\!
-\int \! \frac{d^nl}{i(2 \pi)^n} 
\, \tr \, 
\Big[
\hspace{0mm}
\overbrace{ 
-i\Big(
\gamma_5\, \frac{1}{-({\slash ~ l}+{\slash ~ K_2})} 
+\frac{1}{-{\slash ~ l}}  \, \gamma_5
\Big)
\, \gamma^\rho
\frac{1}{-({\slash ~ l}+{\slash ~ K_1})} 
\, \gamma^\nu}^{\textrm{$\to 0$  as shown in (\ref{wrosv})}}
\nonumber \\*[1.0mm]  
%---         
&&\hspace{25.5mm}
+\frac{1}{-{\slash ~ l}} \,
(-2i {\slash ~ r} \gamma_5 ) \,
\frac{1}{-({\slash ~ l}+{\slash ~K_2})} \gamma^\rho 
\frac{1}{-({\slash ~ l}+{\slash ~K_1})} \gamma^\nu                             
\Big],
\end{eqnarray}
\vspace{-1.0mm}
where $\widetilde{\Gamma}^{(3)}_{\cal A}$ means 
the $n$-dimensional version of $\Gamma^{(3)}_{\cal A} \,(\equiv \Gamma^{\mu\rho\nu}_{\cal A})$. 
Here,
\begin{itemize}
\item
The first two terms vanish at an arbitrary real $n$ as shown in (\ref{wrosv}). 
This is reflecting the BRS symmetry. 
Therefore, that the term irrelevant to ${\slash ~ r}$ vanishes is universal 
in $ik_n{}_\mu \widetilde{\Gamma}_{\cal A}^{(n)}$. 
Along with this, 
that the term proportional to ${\slash ~ r}$ remains (the third term) is also universal 
in $ik_n{}_\mu \widetilde{\Gamma}_{\cal A}^{(n)}$. 
\item
The third term appears to vanish at $n \to 4$ as ${\slash ~ r} \to 0$ at $n \to 4$,
however, it includes the UV-divergence, 
and these vanishment and divergence are multiplied by each other. 
As a result, a coefficient remains as a finite contribution at $n \to 4$.   
\item
From the these two points, it can be seen that 
the source of the finite contribution is $-2 {\slash ~ r}\, \gamma_5$ in (\ref{tksjy}) 
(and the UV-divergence).
$-2 {\slash ~ r}\, \gamma_5$ arises when $\gamma_5$ is included in the addressed current. 
Therefore, the finite contribution arises when the current is the chiral type.  
\end{itemize}

$
(\cdots) \, 
i\gamma^\rho
\frac{1}{-({\slash ~\, l}+{\slash ~K_1})} 
\, i\gamma^\nu
$ 
in (\ref{asuis}) can be written in the whole (\ref{hbdd3}) like the following, 
and vanishes as
\vspace{-3.0mm}
\begin{eqnarray}\label{wrosv}
\!\! && \!\! 
\hspace{2.0mm}
i\int \Big(\prod_{i=1}^3 \frac{d^nk_i}{(i(2\pi)^4)}\Big) \, 
(2\pi)^4 \, 
\delta^4 (K_3) ~
\varepsilon(-k_3)\, c(-k_3) \, A_\nu (-k_1) \, A_\rho (-k_2) 
\nonumber \\*[1.0mm]
%---
&& \hspace{-1.5mm}
\times 
\int \! \frac{d^nl}{i(2 \pi)^n} 
\, \tr \, 
\Big[
\gamma_5 \,
\Big(
 \frac{1}{{\slash ~ l}+{\slash ~ K_2}}\, \gamma^\rho  \,
 \frac{1}{{\slash ~ l}+{\slash ~ K_1}}                \, \gamma^\nu   
- 
\frac{1}{{\slash ~ l}+{\slash ~ K_1}}   \,\gamma^\nu  \,       
 \frac{1}{{\slash ~ l}}                               \,\gamma^\rho\,    
\Big)
\Big]=0.
\end{eqnarray}
This is because, first, note that 
$A_\mu=(A_0,A_1,A_2,A_3, 0,\cdots,0)$;
thus always $\gamma_5 \gamma^\mu=-\gamma_5 \gamma^\mu$ in (\ref{wrosv}).
Then, in the first term, if we exchange $(\rho,k_2)$ and $(\nu,k_1)$ with each other, 
then, shift $l_\mu$ as $l_\mu \to l_\mu-k_2{}_\mu$, 
the first term can coincide with the second term with the opposite sign.

With (\ref{mrrgy1}), (\ref{asuis}) can be given as follows:
\vspace{-1.0mm}
\begin{eqnarray}\label{aavsy} 
\textrm{(\ref{asuis})}
\!\!\! &=& \!\!\!
-2i
\int \! \frac{d^nl}{i(2 \pi)^n} 
\frac
{\tr \,[
{\slash ~ l}
({\slash ~ r} \gamma_5 )
({\slash ~ l}+{\slash ~ K_2})\, \gamma^\rho 
({\slash ~ l}+{\slash ~ K_1})\, \gamma^\nu
]}
{l^2 \, (l+K_2)^2 \, (l+K_1)^2 }
\nonumber\\*[1.0mm]
%---
\!\!\! &=& \!\!\!
-4i
\int_0^1 y\,dy \int_0^1 dx 
\int \! \frac{d^nl}{i(2 \pi)^n} 
\frac
{\tr \,[
(\gamma_5 {\slash ~ r} )
({\slash ~ l}+{\slash ~ K_2})\, \gamma^\rho 
({\slash ~ l}+{\slash ~ K_1})\, \gamma^\nu
{\slash ~ l} \,
]}
{\Omega^3},
\\*[1.0mm]
%---
\Omega  \!\!\! &\equiv&  \!\!\! -l^2+2\Delta \,l+\Sigma^2,\quad
\Delta^\mu  \equiv -(x K_1-(1-x)K_2 )^\mu \,y, \quad
\Sigma^2 \equiv  -(x (K_1)^2+(1-x)(K_2)^2)\,y,
\nonumber
\end{eqnarray} 
where, as for how to take $\Delta^\mu$ and $\Sigma^2$, 
there are alternatives to the one above, 
but none of which affects the final result. 
%---
The divergence-order at (\ref{ttshs}) is $1$ (supposing $n=4$),
however, it has been changed to $2$ as can be seen in (\ref{asuis}) or (\ref{aavsy})
as a result of the implementation of the dimensional regularization. 
The same thing happens in the calculation of (\ref{hbdd5}).

Using (\ref{dweklv}) in the numerator of (\ref{aavsy}), 
it is given to the third-order of ${\slash ~ r}$ maximally as follows:
\vspace{-2.5mm}
\begin{eqnarray}\label{vewr}
\textrm{(\ref{aavsy})}
=
-4i
\int_0^1 y\,dy \int_0^1 dx 
\int \! \frac{d^nl}{i(2 \pi)^n} 
\frac
{\, \tr \,[ 
(\gamma_5 {\slash ~ r} ) \{ 
\overbrace{(\cdots){\slash ~ r}^1}^{\to \,(\ref{aehko})}
+\overbrace{(\cdots){\slash ~ r}^0}^{\to \,(\ref{svoyt})}
+\overbrace{(\cdots){\slash ~ r}^2}^{\to \,(\ref{sutft})} 
+\overbrace{(\cdots){\slash ~ r}^3}^{\to \,(\ref{sebr})} \}
]}
{\Omega^3} \to \textrm{(\ref{klwe})}.
\end{eqnarray}
Each $(\gamma_5 {\slash ~ r} )(\cdots){\slash ~ r}^p$ ($p=0,1,2,3$) includes a product of six $\gamma$-matrices.
Because of 
\begin{eqnarray}\label{reir}
\tr\,[\gamma_5 \gamma^\nu \gamma^\rho \gamma^\alpha \gamma^\beta]=-4i\,\epsilon^{\nu \rho \alpha \beta}, 
\end{eqnarray}
two of them should contract. 
Based on this, we evaluate each term in what follows. 
\begin{itemize}
%=================================
% ITEM 1
%=================================
\item
$\tr \,[(\gamma_5 {\slash ~ r} ) \{ (\cdots){\slash ~ r}^1 \}]$
is concretely given as follows:
\begin{eqnarray}\label{adko}
\hspace{-6mm}
r^2 
\tr [
\gamma_5  
\{ 
\gamma^\rho 
({\slash ~ \uline{l}}+{\slash ~ K_1})\, \gamma^\nu
 {\slash ~ \uline{l}}
+
({\slash ~ \uline{l}}+{\slash ~ K_2})\, \gamma^\rho \, \gamma^\nu
{\slash ~ \uline{l}}
+
({\slash ~ \uline{l}}+{\slash ~ K_2})\, \gamma^\rho 
({\slash ~ \uline{l}}+{\slash ~ K_1})\, \gamma^\nu
\}]
+r_\alpha v^\alpha\, r_\beta \, \tr [\gamma_5  (\cdots)^{\beta \rho \nu}
], 
\end{eqnarray}
where 
\vspace{-2.0mm}
\begin{itemize}
\item
The first term is the contribution by two ${\slash ~ r}$ contracting, 
where ${\slash ~ r}\!{\slash ~ r}=r^2$ 
(the source of these two ${\slash ~ r}$ are that in $(\gamma_5 {\slash ~ r})$ and that in $\{ (\cdots){\slash ~ r}^1 \}$). 
\item
The second term is the not the contribution by two ${\slash ~ r}$ contracting. 
\begin{itemize}
\item
$v^\alpha$ means some vector other than $r^\alpha$, 
which is some of $K_i$ ($i=1,2,3$) or $\uline{l}^\alpha$. 
Therefore, we can denote $v^\alpha$ as $\underline{v}^\alpha$ in the same fashion as $\underline{l}_\mu$ in (\ref{dweklv}).
\item
Since the second term finally vanishes no matter what $v_\alpha \tr [\gamma_5  (\cdots)^{\beta \rho \nu}]$ is (as shown in (\ref{btern})), 
we do not need to concretely present $v^\alpha \tr [\gamma_5  (\cdots)^{\beta \rho \nu}]$.
\end{itemize}
\item
Each term consists of four $\gamma$-matrices 
as a result of two $\gamma$-matrices having contracted. 
\end{itemize}
The first three terms in (\ref{adko}) can be given as follows:
\vspace{-1.0mm}
\begin{subequations}\label{swvnio}
\begin{align}
%-----
\label{swvnio1}
r^2\, \tr \,[ 
\gamma_5 \,\,  
\gamma^\rho 
({\slash ~ \uline{l}}+{\slash ~ K_1})\, \gamma^\nu
{\slash ~ \uline{l}}
\,]
=& \,
r^2\,\tr \, [\gamma_5 \gamma^\rho  \gamma^\alpha \gamma^\nu \gamma^\beta]
\,({\slash ~ \uline{l}}+{\slash ~ K_1})_\alpha \,\uline{l}_\beta,
\\*[1.0mm]
%-----
\label{swvnio2}
r^2\,\tr \,[
\gamma_5 \, 
({\slash ~ \uline{l}}+{\slash ~ K_2})\, \gamma^\rho 
\, \gamma^\nu
{\slash ~ \uline{l}}
\,]
=& \,\, 
r^2\,\tr \, [\gamma_5 \gamma^\alpha  \gamma^\rho \gamma^\nu \gamma^\beta]
\,({\slash ~ \uline{l}}+{\slash ~ K_2})_\alpha \,\uline{l}_\beta, 
\\*[1.0mm]
%-----
\label{swvnio3}
r^2\,\tr \,[ 
\gamma_5 \,
({\slash ~ \uline{l}}+{\slash ~ K_2})\, \gamma^\rho 
({\slash ~ \uline{l}}+{\slash ~ K_1})\, \gamma^\nu
]
=& \,\,
r^2\,\tr \, [\gamma_5 \gamma^\alpha  \gamma^\rho \gamma^\beta \gamma^\nu]
\,({\slash ~ \uline{l}}+{\slash ~ K_2})_\alpha \,({\slash ~ \uline{l}}+{\slash ~ K_1})_\beta. 
\end{align} 
\end{subequations}
Summing up (\ref{swvnio1})-(\ref{swvnio3}), 
\vspace{-1.0mm}
\begin{eqnarray}\label{aehko}
\textrm{(\ref{adko})}
\!\!\! &=& \!\!\!
-r^2 \,
\tr \, [\gamma_5 \gamma^\beta  \gamma^\rho \gamma^\alpha \gamma^\nu] \,
K_{1\beta} K_{2\alpha}
+r_\alpha v^\alpha\, r_\beta \,\tr \,[\gamma_5 (\cdots)],
\end{eqnarray}
where 
$K_{1\beta} K_{2\alpha}$ can be given as $k_{1\beta} k_{2\alpha}$ 
using (\ref{reir}) at $n \to 4$. 

%=================================
% ITEM 2
%=================================
\item
$\tr \,[(\gamma_5 {\slash ~ r} ) \{ (\cdots){\slash ~ r}^0\}]$
is concretely given and vanishes as follows:
\vspace{-1.0mm}
\begin{eqnarray}\label{svoyt}
%============
\int \! \frac{d^nl}{i(2 \pi)^n} \,
\frac{
\tr \,[
(\gamma_5 {\slash ~ r} )
\{
({\slash ~ K_2})\, \gamma^\rho 
({\slash ~ K_1})\, \gamma^\nu
{\slash ~ \uline{l}} \,
\}]
}{\Omega^3}
=
\tr\,[(\gamma_5 \gamma^\alpha)(
{\slash ~ K_2} 
\gamma^\rho  
{\slash ~ K_1} 
\gamma^\nu  
\gamma^\beta)] \,
\int \! \frac{d^nl}{i(2 \pi)^n} \,
\frac{r_\alpha\,\uline{l}_\beta}{\Omega^3}
\xrightarrow [n \to 4]{}0.
\end{eqnarray}
The result of (\ref{svoyt}) can be obtained 
by dividing the middle equation into two parts with 
$r_\alpha\,\uline{l}_\beta=(l_\alpha-l_{\uline{\alpha}}) \, l_{\uline{\beta}}$, 
and then, using the formulae:
\vspace{-4.5mm}
\begin{subequations}\label{ewsiu} 
\begin{align} 
\label{ewsiu1} 
\int \! \frac{d^nk}{i(2\pi)^n}\frac{k_\alpha \uline{k}_\beta}{{\cal B}^3}
=& \,\,
\frac{1}{(4\pi)^\eta \,\Gamma(3)}\,
\Big(
\overbrace{\Gamma(3-\eta) \, 
\frac{p_\alpha \uline{p}_\beta}{{\cal C}^{3-\eta}}}^{\rm finite}
-\overbrace{\Gamma(2-\eta) \, 
\frac{\uline{g}_{\alpha\beta}}{2{\cal C}^{2-\eta}}}^{\rm divergent}
\Big), 
\quad \eta \equiv \frac{n}{2},
\\*[1.0mm]
\label{ewsiu2} 
\int \! \frac{d^nk}{i(2\pi)^n}\frac{\uline{k}_\alpha \uline{k}_\beta}{{\cal B}^3}
=& \,\,
\frac{1}{(4\pi)^\eta \,\Gamma(3)}\,
\Big(
\Gamma(3-\eta) \, 
\frac{\uline{p}_\alpha \uline{p}_{\beta}}{{\cal C}^{3-\eta}}
-\Gamma(2-\eta) \, 
\frac{\uline{g}_{\alpha\beta}}{2{\cal C}^{2-\eta}}
\Big),
\end{align} 
\end{subequations}
for these parts, respectively.  
These were obtained from (\ref{mrrgy5})
by multiplying (\ref{mrrgy5}) by 
$\theta_{\mu\alpha} \uline{\theta}_{\nu\beta}$ and
$\uline{\theta}_{\mu\alpha} \uline{\theta}_{\nu\beta}$, respectively 
($\uline{\theta}_{\alpha\beta} \equiv \diag_{\textrm{$n$-dim.}}(1,1,1,1,0,\cdots,0)$). 
$\uline{p}_\alpha \equiv \uline{\theta}_{\alpha \beta} p^\beta$ and
$\uline{g}_\alpha{}^\beta \equiv \uline{\theta}_{\alpha\gamma} g^{\gamma\beta}$. 
($\theta_{\gamma\beta}\,\uline{g}_\alpha{}^\beta=\uline{\theta}_{\gamma\beta}\,\uline{g}_\alpha{}^\beta=\uline{g}_{\alpha\gamma}$).  
\begin{itemize}
\item 
The finite parts in the $l_\alpha l_{\uline{\beta}}$ and $l_{\uline{\alpha}}l_{\uline{\beta}}$ terms 
always cancel each other out at $n \to 4$ as $p_{\alpha} \xrightarrow [n \to 4]{} \uline{p}_{\alpha}$. 
This is understandable as these are originally the terms in the term proportional to ${\slash ~ r}$ in (\ref{asuis}).
\item 
In the divergent part, $\Gamma(2-\eta)$ is diverged at $n \to 4$.
However, in the case that the numerator does not contract, 
the numerator is 0, and there is no contribution for arbitrary $n$. 
On the other hand, even in the case that the numerator contracts, 
the divergent parts in (\ref{ewsiu1}) and (\ref{ewsiu2}) are the same as each other,
and its difference is 0 for arbitrary $n$.

\item
The existence of the divergent part is caused in the UV-divergence of (\ref{hbdd3}).
Conversely, if the UV-divergence were not in (\ref{hbdd3}), such a divergent part would not appear. 
Therefore, such divergent part does not appear in (\ref{hbdd6}). 
\end{itemize}

%=================================
% ITEM 3
%=================================
\item
$\tr \,[(\gamma_5 {\slash ~ r} ) \{ (\cdots){\slash ~ r}^2\}]$ 
exists totally ${}_3C_2=3$ patterns, 
and vanishes at $n \to 4$ as follows:
\vspace{-1.0mm}
\begin{subequations}\label{sutft}
\begin{align}
\label{sutft1}
&\hspace{+9.5mm}
\int \! \frac{d^nl}{i(2 \pi)^n} \,
\frac{
\tr\,[(\gamma_5 {\slash ~ r} )
({\slash ~ r})\, \gamma^\rho 
({\slash ~ r})\, \gamma^\nu
 {\slash ~ \uline{l}}\,]}
 {\Omega^3}
=
\tr\,[\gamma_5 \gamma^\rho  \gamma^\alpha \gamma^\nu \gamma^\beta] \,
\int \! \frac{d^nl}{i(2 \pi)^n} \,\frac{r^2 \, r_\alpha \uline{l}_\beta}{\Omega^3}
\xrightarrow[n \to 4]{}0,
\\*[1.0mm]
%----------
\label{sutft2}
&\hspace{-2mm}
\int \! \frac{d^nl}{i(2 \pi)^n} \,
\frac{
\tr\,[(\gamma_5 {\slash ~ r} )
({\slash ~ r})\, \gamma^\rho 
({\slash ~ \uline{l}}+{\slash ~ K_1})\, \gamma^\nu
 {\slash ~ r}\,]}
 {\Omega^3}
=
\tr\,[\gamma_5 \gamma^\rho  \gamma^\alpha \gamma^\nu \gamma^\beta] \,
\int \! \frac{d^nl}{i(2 \pi)^n}
\frac{r^2 \,(\uline{l}+K_1)_\alpha \, r_\beta
}{\Omega^3} 
\xrightarrow[n \to 4]{}0,
\\*[1.0mm]
%----------
\label{sutft3}
&\hspace{-2mm}
\int \! \frac{d^nl}{i(2 \pi)^n} \,
\frac{
\tr\,[(\gamma_5 {\slash ~ r} )
({\slash ~ \uline{l}}+{\slash ~ K_2})\, \gamma^\rho 
({\slash ~ r})\, \gamma^\nu
 {\slash ~ r}\,]}
{\Omega^3}
=
\tr\,[\gamma_5 \gamma^\alpha  \gamma^\rho \gamma^\nu \gamma^\beta] \,
\int \! \frac{d^nl}{i(2 \pi)^n}
\frac{r^2 \,(\uline{l}+K_2)_\alpha \, r_\beta
}{\Omega^3} 
\xrightarrow[n \to 4]{}0,
\end{align} 
\end{subequations}
where 
$
r^2 r_\alpha\uline{l}_\beta
=
l^2(l_\alpha\uline{l}_\beta-{\uline{l}}_\alpha\uline{l}_\beta)
-\uline{l}^2(l_\alpha\uline{l}_\beta-{\uline{l}}_\alpha\uline{l}_\beta)
$ and 
$
r^2 r_\beta
=
l^2 (l-{\uline{l}})_\beta
-\uline{l}^2 (l-{\uline{l}})_\beta
$, and the results of (\ref{sutft1})-(\ref{sutft3}) can be obtained with the following formulae 
obtained from (\ref{mrrgy3}) and (\ref{mrrgy4}):
\vspace{-1.0mm}
\begin{subequations}\label{dtjve}
\begin{align}
\label{dtjve1}
\int \! \frac{d^nk}{i(2\pi)^n}\frac{k^2 k_\alpha \uline{k}_\beta}{{\cal B}^3}
=& \,
\frac{1}{{\cal D}}
\Big(
\Gamma(3-\eta) \, \frac{p^2 \, \uline{p}_\alpha \uline{p}_\beta}{{\cal C}^{3-\eta}}
-\Gamma(2-\eta) \frac{ 
 (n+4) \, \uline{p}_\alpha \uline{p}_\beta
+ p^2 \,\uline{g}_{\alpha \beta} }{2\,{\cal C}^{2-\eta}}
+\Gamma(1-\eta) \frac{(n+2)\, \uline{g}_{\alpha \beta} 
}{4\,{\cal C}^{1-\eta}}
\Big), 
\\*[-2.0mm]
%----------
\label{dtjve3}
\int \! \frac{d^nk}{i(2\pi)^n}\frac{k^2 k_{\uline{\alpha}}}{{\cal B}^3}
=& \,
\frac{1}{{\cal D}}
\Big(  
\Gamma(3-\eta)\,\frac{p^2 \, \uline{p}_\alpha}{{\cal C}^{3-\eta}}
-\Gamma(2-\eta) 
\frac{(n+2)\,\uline{p}_\alpha
}{2\,{\cal C}^{2-\eta}}
\Big),
\end{align}
\end{subequations}
where $p_\alpha \uline{p}_\beta=\uline{p}_\alpha \uline{p}_\beta$ 
(we skip giving other necessary formulae).

%=================================
% ITEM 4
%=================================
\item
$\tr \,[(\gamma_5 {\slash ~ r} ) \{ (\cdots){\slash ~ r}^3\}]$
is concretely given as follows and can be known to vanish as 
\vspace{-1.0mm}
\begin{eqnarray}\label{sebr}
\tr\,[
(\gamma_5 {\slash ~ r} )
({\slash ~ r})\, \gamma^\rho 
({\slash ~ r})\, \gamma^\nu
 {\slash ~ r}\,
]
= 
\tr\,[(\gamma_5)\,\gamma^\rho \,\gamma^\nu ] \,(r^2)^2=0.
%----------
\end{eqnarray} 
\end{itemize}

\vspace{-1.0mm}
Assembling the results of 
(\ref{aehko}), (\ref{svoyt}), (\ref{sutft}) and (\ref{sebr}), 
$ik_3{}_\mu \widetilde{\Gamma}_{\cal A}^{(3)}$ in (\ref{vewr}) can be given as follows:
\vspace{-1.0mm}
\begin{eqnarray}\label{klwe}
ik_3{}_\mu \widetilde{\Gamma}_{\cal A}^{(3)} 
=
-4i\int_0^1 \! y\,dy \int_0^1 \! dx 
\int \! \frac{d^nl}{i(2 \pi)^n\,\Omega^3} \Big(\!
-\tr\,[\gamma_5 \gamma^\beta  \gamma^\rho \gamma^\alpha \gamma^\nu] \, 
k_{1\beta} k_{2\alpha}
\,(l^2-\uline{l}^2)
+
\tr \,[\gamma_5 (\cdots)] \,
v^\alpha \, r_\alpha  r_\beta \Big).
\end{eqnarray}
The integrals in the one above result in the following:
\vspace{-1.0mm}
\begin{subequations}\label{btern}
\begin{align} 
%-----
\int_0^1 \! y\,dy \int_0^1 \! dx \int \frac{d^nl}{i(2 \pi)^n} 
\frac{l^2-\uline{l}^2}{\Omega^3}
=& \,
\int_0^1 \! y\,dy \int_0^1 \! dx \,
\frac{1}{(4\pi)^\eta \, \Gamma(3)} 
\Big(
 \frac{\Gamma(3-\eta) (p^2-\uline{p}^2)}{(\Delta^2+\Sigma^2)^{3-\eta}}
-\frac{\Gamma(2-\eta)(n-4)}{2(\Delta^2+\Sigma^2)^{2-\eta}}
\Big)
\nonumber
\\*[1.0mm]
\label{btern1}
\xrightarrow[n \to 4]{}& \,
\int_0^1 \! y\,dy \int_0^1 \! dx \, \frac{1}{32\pi^2}=\frac{1}{64\pi^2}, 
\\*[1.0mm]
\label{btern2}
&\hspace{-60mm}
\int \! \frac{d^nl}{i(2 \pi)^n} \,
\frac{v^\alpha \, r_\alpha  r_\beta }{\Omega^3} 
= \,
-\frac{\Gamma(2-\eta)}{(4\pi)^\eta \, \Gamma(3)} 
\frac{v^\alpha(g_{\alpha\beta} - \uline{g}_{\alpha\beta})}
{2(\Delta^2+\Sigma^2)^{2-\eta}} 
=
-\frac{\Gamma(2-\eta)}{(4\pi)^\eta \, \Gamma(3)} 
\frac{v_{\beta} - \uline{v}_{\beta}}
{2(\Delta^2+\Sigma^2)^{2-\eta}} =0.
\end{align} 
\end{subequations}
(\ref{btern1}) can be known from (\ref{ewsiu}). 
$n=g^\mu{}_\mu$ and $4=\uline{g}^\mu{}_\mu$, and $n-4$ converges to 0 at $n \to 4$, 
which cancels the divergence from $\Gamma(2-\eta)$, and its coefficient remains.
In (\ref{btern2}), 
since $v_{\beta}$ can be denoted as $\uline{v}_{\beta}$ as mentioned under (\ref{adko}), 
$v_{\beta} - \uline{v}_{\beta}$ vanishes for arbitrary $n$.
With (\ref{reir}), $ik_\mu \Gamma_{\cal A}^{(3)}$ finally results in 
\vspace{-1.0mm}
\begin{eqnarray}\label{jydva}
ik_{3\mu} \Gamma_{\cal A}^{(3)} = \epsilon^{\beta \rho \alpha \nu} k_{1\beta} k_{2\alpha}/4 \pi^2.
\end{eqnarray}

\vspace{-4.0mm}
%=============================================
\subsection{The calculations of terms other than (\ref{hbdd3})}
\label{hbddT4}
%=============================================  

We turn to the terms other than (\ref{hbdd3}), 
which are (\ref{hbdd1}), (\ref{hbdd2}), (\ref{hbdd4}), (\ref{hbdd5}) and (\ref{hbdd6}).
\begin{itemize}
\item
It can be seen immediately that (\ref{hbdd1}) vanishes 
because of $\delta^4(k_1) \, k_{1\mu}$ in that equation.
%-----
\item
As for (\ref{hbdd2}), 
its calculation can proceed
in the same way as that in Sec.\ref{hbrbes}, 
and the following $l$-integral appears at the stage corresponding to (\ref{aavsy}):
\vspace{-1.5mm}
\begin{eqnarray}\label{ebmry} 
\int \! \frac{d^nl}{i(2 \pi)^n} 
\frac
{\tr \,[ 
{\slash ~ l}
({\slash ~ r} \gamma_5)
({\slash ~ l}+{\slash ~ K_1})\, \gamma^\rho 
]}
{l^2 \, (l+K_1)^2}
=
\int \! \frac{d^nl}{i(2 \pi)^n} 
\frac
{\tr \,[
\hspace{-30mm}
\overbrace{{\slash ~ l}\,({\slash ~ r} \gamma_5){\slash ~ l}\, \gamma^\rho}^{
\textrm{Four $\gamma$-matrices, which becomes less than four after contractions}} 
\hspace{-31mm}
+{\slash ~ l}\,({\slash ~ r} \gamma_5){\slash ~ K_1}\, \gamma^\rho 
]}
{l^2 \, (l+K_1)^2}.
\end{eqnarray}
${\slash ~ r}$ is replaced with (\ref{dweklv}). 
Then, in order for $l$-integral to yield finite contributions at $n \to 4$, 
all ${\slash ~ l}$ and ${\slash ~ \uline{l}}$ should contract in its numerator 
(for this, see the points under (\ref{ewsiu})). 
As a result, the number of $\gamma$-matrices becomes less than four in each term.
However, four $\gamma$-matrices are needed for $\tr[\gamma_5\cdots]$ to remain finitely.
From these, it can be seen that (\ref{ebmry}) vanishes at $n \to 4$.

%-----
\item 
The calculation of (\ref{hbdd4}) can proceed
in the same way as that in Sec.\ref{hbrbes}, and it finally vanishes at $n \to 4$.
This can be seen from the equation corresponding to (\ref{aavsy}):
\vspace{-1.0mm}
\begin{eqnarray}\label{pyuo}
\int\! \Big( \prod_{i=1}^4 \frac{d^4k_i}{i(2\pi)^4} \Big)
(2\pi)^4 \, \delta^4 ( K_4 ) \,
\Big( 
c(-k_4) \, 
\prod_{i=1}^3 A_{\mu_{i}}(-k_{i}) 
\Big) 
\int \! \frac{d^4l}{i(2\pi)^4} \, 
\frac
{
\hspace{-11mm}
\overbrace{
\tr \,[
{\slash ~ l} \,
({\slash ~ r} \gamma_5) \,
\prod_{i=3}^1
({\slash ~ l}+{\slash ~ K_i}) \,\gamma^{\mu_i} 
]}^{\textrm{
Four $\gamma$-matrices are needed for this to remain}}\hspace{-10mm}}  
{l^2(l+K_3)^2(l+K_2)^2(l+K_1)^2}. 
\end{eqnarray}
As mentioned above, four $\gamma$-matrices are needed, 
meanwhile, three $\gamma$-matrices have  already appeared as
\vspace{-6.0mm}
\begin{eqnarray}\label{pwrt}
&&
\tr \,[
{\slash ~ l}\,
({\slash ~ r} \gamma_5) \,
\prod_{i=3}^1
({\slash ~ l}+{\slash ~ K_i}) 
\hspace{-7.0mm}
\overbrace{\gamma^{\mu_i}}^{\textrm{Three $\gamma$-matrices}
} 
\hspace{-7.0mm}
].
\end{eqnarray}
Therefore, one $\gamma$-matrix should appear from the remaining part by some contractions.
From this, the final form will be as 
$v_{\mu_4} \prod_{i=1}^3 A_{\mu_{i}}(-k_{i})\,\epsilon^{\mu_4 \mu_3 \mu_2 \mu_1}$ 
($v_{\mu_4}$ is $k_{i\mu}$, $(k_{i}+k_{j})_\mu$ or $(k_{i}+k_{j}+k_{k})_\mu$).
Then, by exchanging $(\mu_i, k_i) \leftrightarrow (\mu_j, k_j)$  ($i,j =$ $1,2$ or $3$), 
this can be seen to vanish.

\item
From the counting of the divergent-order, the term of (\ref{hbdd5}) can be regarded as non-divergent.
However, according to the situation mentioned under (\ref{aavsy}), there is some possibility that it includes divergence.
However, in (\ref{hbdd5}), four $\gamma$-matrices have already appeared, 
and the form which will be finally realized is $\prod_{i=1}^4 A_{\mu_{i}}(-k_{i})\,\epsilon^{\mu_4 \mu_3 \mu_2 \mu_1}$. 
this can be seen to vanish with the exchange of $(\mu_i, k_i) \leftrightarrow (\mu_j, k_j)$.

\item
In the $n$-dimensional calculation for each term in (\ref{hbdd6}), 
at the stage corresponding to (\ref{asuis}), 
the terms irrelevant to ${\slash ~ r}$ vanish, 
and the term proportional to ${\slash ~ r}$ remains, as well as (\ref{asuis}). 
Since terms in (\ref{hbdd6}) are not divergent, 
that ${\slash ~ r}$-proportional term yields only finite contributions, and vanishes at $n \to 4$ as ${\slash ~ r} = 0$ at $n \to 4$.
\end{itemize}

\vspace{-2.0mm}  
%=============================================
\subsection{The result of (\ref{hbdd})}
\label{hbddT5}
%=============================================  
 
With (\ref{jydva}) and the results in Sec.\,\ref{hbddT4}, 
(\ref{hbdd}) is finally given as follows:
\vspace{-1.0mm}  
\begin{eqnarray} \label{ypfui}
&&
\boldsymbol{\delta}_{\rm \varepsilon} \Gamma_{\cal A}[A]
=
i\int \Big( \prod_{i=1}^3\frac{d^4k_i}{(2\pi)^4} \Big)
(2\pi)^4 \, \delta^4 (K_3) \,
\varepsilon (-k_3) \,c(-k_3)\,  A_{\rho}(-k_1)A_{\nu}(-k_2)
\,\frac{e^2\epsilon^{\beta \rho \alpha \nu} k_{1\beta} k_{2\alpha}}{4 \pi^2}
\nonumber\\*[1.0mm]
%----- 
&=& \!\!\!
-\int d^4x \, \varepsilon(x)c(x) \,\frac{ie^2\epsilon^{\beta \rho \alpha \nu}}{4 \pi^2} 
\partial_\beta  A_{\rho}(x) 
\partial_\alpha A_{\nu}(x)
=
-\int d^4x \, \varepsilon(x) c(x) \,\frac{ie^2\epsilon^{\beta \rho \alpha \nu}}{16 \pi^2} 
F_{\beta\rho}(x)F_{\alpha\nu}(x),
\end{eqnarray}
where
$A_\mu (k) = \int d^4k e^{ikx} A_\mu (x)$ and 
$\varepsilon (k) c(k) = \int d^4k e^{ikx} \varepsilon (x)c (x)$, 
the inverse of these in Sec.\,\ref{ypvsb2},
and $ \int d^4k e^{-ikx} = (2\pi)^4\delta^4(x)$.
Applying this result to (\ref{efgnds}), 
the identity agreeing with (\ref{reois}) can be obtained:
\vspace{-1.0mm}
\begin{eqnarray}\label{srds}
\varepsilon \,i\,\frac{e^3}{16\pi^2}\epsilon^{\mu\nu\rho\sigma} F_{\mu\nu}F_{\rho\sigma}
=
\partial_\mu \varepsilon\, \langle J^\mu_{c} \rangle
-\partial_\mu \langle J^\mu_{c \epsilon} \rangle.
\end{eqnarray}

\vspace{-5.0mm} 
%=============================================
\section{The anomaly equation by the Fujikawa method}
\label{itioh}
%=============================================  

The anomaly equation associated with the large chiral symmetry was 
obtained from the heuristically constructed large chiral charge in Sec.\,\ref{ykiwf}
and the evaluation of the BRS-transformed one-loop diagrams in  Sec.\,\ref{ypvsb}.
In this section, we derive it from the Fujikawa method.
\newline

Under the large chiral transformation (\ref{zery}) in the Cartesian coordinates, 
the variations of action (\ref{erer}) and the measure of the Dirac field are given as 
\vspace{-1.0mm}
\begin{subequations}
\begin{align}
\label{tdvul1}
&\delta S_0
 =  \int d^4x \,i(\bar{\psi}'\!\slash{D}\psi'-\bar{\psi}\!\slash{D}\psi)
 = e\int d^4x \, \partial_\mu(\lambda\varepsilon) \, \bar{\psi}\gamma^\mu \gamma_5\psi
 =  \int d^4x \,\lambda\,( \partial_\mu(\varepsilon J_c^\mu) - (\partial_\mu \varepsilon) J_c^\mu),
\\*[0.5mm]
%---
\label{tdvul2}
&{\cal D \bar{\psi}'}{\cal D}\psi' 
= 
\exp\Big[\!-2i\int \! d^4x \,e\lambda\varepsilon\,
{\cal A}
\Big] 
{\cal D \bar{\psi}}{\cal D}\psi,
\quad
%---
{\cal A}
\equiv
\lim_{\Lambda \to \infty}
(\sum_{n=-\infty}^\infty \varphi^\dagger_n \,\gamma_5 \,e^{-(\eta_n/\Lambda)^2}\varphi_n)
=
-\frac{e^2}{32\pi^2}\,\epsilon^{\mu\nu\rho\sigma}\,F_{\mu\nu}F_{\rho\sigma},
\end{align}
\end{subequations}
where $\lambda$ has been taken to the linear order and is supposed to vanish at $r \to \infty$. 
$ J_c^\mu$ and $ J_{c\varepsilon}^\mu$ are defined in (\ref{gxsrj}).
$\varphi_n$ and $\varphi_n^\dagger$ are the eigenvectors as ${\slash D} \varphi_n=\eta_n \varphi_n$, 
and form a complete orthogonal set: 
$\int d^4x \,\varphi_m^\dagger \varphi_n=\delta_{mn}$.
With these, $\psi$ and $\bar{\psi}$ have been expanded.
$e^{-(\eta_n/\Lambda)^2}$ have the cut-off role for $\eta_n \gg \Lambda$.
The calculation  of ${\cal A}$ in (\ref{tdvul2}) is well-known; 
thus we have noted the results. 

Giving $Z_{\psi,A}\equiv \int {\cal D}\bar{\psi}{\cal D}\psi{\cal D}A \, e^{iS_0}$
with (\ref{tdvul1}) and (\ref{tdvul2}),
\vspace{-1.0mm}  
\begin{eqnarray}\label{efwe}
Z_{\psi',A'} \! = \!
\int {\cal D \bar{\psi}}{\cal D}\psi{\cal D}A 
\exp\Big[\int \! d^4x 
\,\Big\{
{\cal L}_0
+ 
\lambda
\Big( 
\partial_\mu (\varepsilon J_c^\mu) 
- (\partial_\mu\varepsilon) J_c^\mu 
+\varepsilon\,\frac{i e^3 }{16\pi^3} \,\epsilon^{\mu\nu\rho\sigma}\,F_{\mu\nu}F_{\rho\sigma}
\Big) \!
\Big\}
\Big].
\end{eqnarray}
From the invariance $Z_{\psi',A'}=Z_{\psi,A}$, 
the following anomaly equation can be obtained:  
\vspace{-1.0mm}
\begin{eqnarray}\label{tldgd}
 \partial_\mu \langle J_{c\varepsilon}^\mu \rangle 
-\partial_\mu\varepsilon \,\langle J_c^\mu \rangle 
+\varepsilon \,\frac{ie^3}{16\pi^2}\,\epsilon^{\mu\nu\rho\sigma}\,F_{\mu\nu}F_{\rho\sigma}=0,
\end{eqnarray}
where $\varepsilon J_c^\mu \equiv J_{c\varepsilon}^\mu$ as seen in (\ref{gxsrj}). 
We can see this agrees with (\ref{reois}) and (\ref{srds}).

Taking the limit of $\Lambda$ prior to the summation of $n$ in the middle equation of ${\cal A}$,  
it is given as $\sum_n \varphi^\dagger_n \,\gamma_5 \,\varphi_n$.
%---
Then, ${\slash D}\, \gamma_5 \varphi_n=-\eta_n \gamma_5\varphi_n$,
thus, $\gamma_5 \varphi_n$ are the eigenvectors associated with the eigenvalues $-\lambda_n$, 
and $\varphi_n$ and $\gamma_5 \varphi_n$ are orthogonal to each other. 
From this, $\int d^4x \, {\cal A}$ results in $n_+-n_-$ 
($n_\pm$ mean the number of the zero-mode particles with each positive/negative chirality, respectively)\footnote{
%====================
%===== FOOTNOTE =====
%====================
Since ${\slash D}$ and $\gamma_5$ can be simultaneously diagonalized, 
$
\int d^4x \, {\cal A} 
= \sum_n \int d^4x \, ( \varphi^{(+)\dagger}_n \,\varphi_n^{(+)} -\varphi^{(-)\dagger}_n \,\varphi_n^{(-)})
= \int d^4x \, ( \varphi^{(+)\dagger}_0 \,\varphi_0^{(+)} -\varphi^{(-)\dagger}_0 \,\varphi_0^{(-)})
$, 
where $\gamma_5\varphi_n=\pm \varphi_n^{(\pm)}$, 
and by the orthogonality between $\varphi_n$ and $\gamma_5 \varphi_n$, 
contributions for $n\not=0$ vanish. 
%====================
%====================
%====================
}.

\vspace{-1.0mm}
%=============================================
\section{The issues related to this study and the future work}
\label{psent}
%=============================================  

In this section, we discuss two issues related to the large chiral anomalies in this study
and comment on potential future developments based on them.

\vspace{-1.0mm}
%=============================================
\subsection{The large chiral anomalies in this study and the breaking of unitarity}
\label{psent1}
%=============================================  

In this subsection, as an issue related to the large chiral anomalies in this study, 
we discuss that 
these large chiral anomalies are 
the quantities of the breaking in the Ward-Takahashi (WT) identity 
and these link to the violation of unitarity.
\newline

First, we denote the large chiral anomalies in this study (\ref{ypfui}) as 
\begin{eqnarray}\label{jnln}
\mathfrak{a}_\varepsilon = 
-\varepsilon(x) c(x) \,\frac{ie^2\epsilon^{\beta \rho \alpha \nu}}{4 \pi^2} 
\partial_\beta A_\rho(x)\partial_\alpha A_\nu(x)
\end{eqnarray}
The BRS transformation of $\mathfrak{a}_\varepsilon$ can be expressed 
in the form of a total derivative as follows:
\begin{eqnarray}\label{gvdy}
\bdelta_{\rm B} \mathfrak{a}_\varepsilon 
=-\varepsilon(x)  \,\frac{i^2e^3\epsilon^{\beta \rho \alpha \nu}}{4 \pi^2}
\partial_\beta (c^2(x)(A_\rho(x)\partial_\alpha A_\nu(x)),
\end{eqnarray}
where this vanishes 
as $c^2(x)=0$ identically vanishes in this study with U(1) gauge field. 
We calculated the one above with $\bdelta_{\rm B} A_\mu = e^{-1}D_\mu c$ 
(and $\bdelta_{\rm B} c =ie c$), 
where $ D_\mu c=\partial_\mu c$ in this study.
Here, the BRS transformation of $A_\mu$ is given in 
(\ref{fytou}) and the covariant derivative is given under (\ref{erer}).
%---

Then, the quantity given by the spacetime integral of $\mathfrak{a}_\varepsilon$, 
which we denote below as $\varDelta_\varepsilon$: 
\begin{eqnarray}\label{gsbs}
\bdelta_{\rm B} \varDelta_\varepsilon  =0, 
\quad \varDelta_\varepsilon \equiv \int d^4x \,\mathfrak{a}_\varepsilon.
\end{eqnarray}
is classified as a non-trivial solution of the Wess-Zumino consistency condition 
$\mathfrak{a}^{\textrm{(non-trivial)}}$
(see Appendix\,\ref{app:stod2} and \ref{app:stod3} 
for the WZ condition and its non-trivial solution, respectively).
\newline

By analogy with the case of the BRS transformation, 
we can see that 
the existence of such a non-trivial $\mathfrak{a}_\varepsilon$
indicates that WT identity regarding the large BRS symmetry is broken for $\mathfrak{a}_\varepsilon$.
%--- 
Its cause is the modification brought about 
by the regularization of the UV-divergence 
(for this point, see under (\ref{svrbe})).
%---
In fact, in Sec.\,\ref{hbrbes},
we employed the dimensional regularization to regularize the UV-divergence. 
%---
As a result, 
a modification 
to treat the $4$ of the four-dimension as a general complex number $n$ 
was implemented to the theory,
and the term $-2 {\slash ~ r}\, \gamma_5$ emerged in (\ref{tksjy}). 
%---
This term is specific to the dimensional regularization and 
ultimately gave rise to the anomalous term in (\ref{jydva}), 
as mentioned under (\ref{asuis}).

$\varDelta_\varepsilon$ itself is merely a solution of the differential equation;
thus, the value of $\mathfrak{a}_\varepsilon$ represents only a possible form of the anomaly. 
%---
In addition, since $\mathfrak{a}_\varepsilon$ is categorized as the non-trivial solution, 
we can see that it cannot be eliminated  regardless of how we modify the regularization scheme we employ.
%---
Thus, by analogy with the logic used in the case of the BRS transformation,
we can conclude that 
the anomalous term in (\ref{jydva}) will always appear in this form  
(for the points in this paragraph, see under (\ref{briu0})).

One problem arising from the breaking of the WT identity 
is that ghost fields appear in the final states.
%---
In fact, 
%---
the BRS symmetry is essential in proving that ghost fields do not appear in the final states, 
and its violation allows ghost fields  to appear in the final states.
Thus, from the similarity between the BRS symmetry and the large BRS symmetry, 
we can infer that the breaking of the large BRS symmetry implies the appearance of ghost fields  in the final states as well.
%---
If ghost fields appear in the final states, unitarity of the system is violated.
Therefore, the large chiral anomalies in this study are linked to the violation of unitarity.

\vspace{-1.0mm}
%=============================================
\subsection{The low-energy effective model}
\label{psent2}
%=============================================

In this subsection, 
as another issue related to the large chiral anomalies in this study, 
we discuss a model 
made up of the U(1) gauge field and the Nambu-Goldstone (NG) particle,
and can reproduce the large chiral anomaly in (\ref{efwe}) 
under its large BRS transformation. 
%-----
This model can be regarded as a low-energy effective model 
for the U(1) gauge symmetric model with a  fermion 
whose $\textrm{U}_{\rm L}(1) \times \textrm{U}_{\rm R}(1)$ chiral symmetry is broken; 
therefore, the NG particle is associated with that broken chiral symmetry. 
Considering this effective model is meaningful. 
%-----

This is because,
for the SU$_\textrm{c}$(3) QCD model 
with a fermion whose $\textrm{SU}_{\rm L}(3) \times \textrm{SU}_{\rm R}(3)$ chiral symmetry is broken, 
a low-energy effective model can be constructed 
by following the condition that 
the anomaly arises in its BRS transformation.
The NG particles in this effective model are interpreted as the pseudoscalar mesons of the octet.
%-----
The decay-width of $\pi \to 2 \gamma$, etc 
can be calculated with the interaction terms obtained from the effective model,
which is in very good agreement with the observational data.
%-----
Therefore, the effective model of the SU$_\textrm{c}$(3) QCD model is important 
in the sense that 
a model constructed on the basis of a theoretical consequence, 
anomaly, agrees with observational results.
%-----
The effective model we consider in this subsection is essentially the same type with this SU$_\textrm{c}$(3) effective model
apart from the difference in their symmetry groups.
\newline

First, 
following the generally known nonlinear representation of the Lagrangian of the NG particle,
we consider the following Lagrangian for $\pi$:
\begin{eqnarray}\label{tyjyr}
{\cal L}_\pi[g]
= -f_\pi^2 (g^{-1} D_\mu g \cdot g^{-1} D^\mu g)
=  f_\pi^2 D_\mu g^{-1} \cdot D^\mu g, \quad 
g(x) \equiv e^{i\pi(x) f_{\pi}},
\end{eqnarray}
where $f_\pi$ is the decay constant.
With this, we consider the following action 
as the effective action we aim at in this subsection: 
\begin{eqnarray}\label{rykr}
\Gamma_{\rm H}[g,A]=\int d^4x \,({\cal L}_\pi-F_{\mu\nu}F^{\mu\nu}/4),
\end{eqnarray}
where 
\begin{itemize}
\item
The index means homogeneous.
Actually, this is invariant under the large BRS transformation (\ref{fytou}): $\bdelta_\varepsilon \Gamma_{\rm H}=0$, 
and the anomaly in (\ref{ypfui}) does not arise. 
Therefore, $\Gamma_{\rm H}$ can be regarded 
as the homogeneous solution of the equation (\ref{ypfui}). 

Generally, the solution of a deferential equation is given by 
``homogeneous solution+particular solution''. 
Therefore, if $\Gamma$ in (\ref{ypfui}) is the solution, 
$\Gamma$ would be presented in some way like $\Gamma_{\rm S}= \Gamma_{\rm H}+\Gamma_{\rm P}$ 
by taking the initial of each. 
We will obtain $\Gamma_{\rm P}$ 
in the following.
\item
The argument $g$ is not included in the r.h.s. of (\ref{rykr}), 
It will appear in $\Gamma_{\rm P}=\Gamma_{\rm P}[g,A]$.
\item
We suppose $\pi$ is massless. 
Actually, in the low-energy effective model for 
the SU$_\textrm{c}$(3) QCD model, 
($u$,$d$,$s$) quarks are approximately regarded as massless, 
along with which, the baryons of the pseudoscalar boson octet are also regarded as massless.
\end{itemize}
~\newline
\vspace{-7.5mm}

As the particular solution mentioned above, 
based on (\ref{ypfui}), 
we consider $\Gamma_{\rm P}$, 
which can produce the large chiral anomaly in this study 
for the large BRS transformation as follows\footnote{
%====================
%===== FOOTNOTE =====
%====================
(\ref{fver}) can be equally presented as 
$-\frac{ie^2}{24\pi^2}\int \omega^1_4(\varepsilon c,A)$, 
where $\omega_{2n+1}(A+C_1,F)=\omega_{2n+1}^0+\omega_{2n}^1+\omega_{2n-1}^2+\cdots+\omega_{0}^{2n+1}$;
$\omega_{2n+1}(A,F)$ is a Chern-Simons form.
However, explaining these requires knowledge of the superfield formalism etc, 
and providing explanations for these would be a major undertaking.  
In addition, the discussion in this study can be performed 
without employing these formalisms.
Therefore, we will not perform the description with this expression in this subsection.
%====================
%====================
%====================
}:
\begin{eqnarray}\label{fver}
\bdelta_{\varepsilon} \Gamma_{\rm P}
= 
-\frac{ie^2\epsilon^{\beta \rho \alpha \nu}}{4\pi^2}
\int d^4x \,\varepsilon(\theta) \, c(x) \, 
\partial_\beta  A_{\rho}(x)
\partial_\alpha A_{\nu}(x),
\end{eqnarray}
where $\theta$ in $\varepsilon(\theta)$ presents the coordinates in the angle direction, 
while $x$ presents all coordinates including the angle direction. 
The large BRS transformation in the one above associates with the following large gauge transformation:
\begin{eqnarray}\label{yjpr}
\delta_\varepsilon g=ie  \,\varepsilon \lambda \, g, \quad 
\delta_\varepsilon A_\mu =e^{-1}\partial_\mu (\varepsilon \lambda),
\end{eqnarray}
where the transformation regarding the gauge field is that in (\ref{ukuy}), 
and $\lambda$ and $\varepsilon$ are those in (\ref{ukuy}).
Then, by performing the reverse procedure of deriving the BRS transformation from the gauge transformation,
we can give (\ref{fver}) as the one for the large gauge transformation as
\begin{eqnarray}\label{mtth}
\delta_\varepsilon \Gamma_{\rm P} 
=
-\frac{ie^2\epsilon^{\beta \rho \alpha \nu}}{4\pi^2}
\int d^4x \,\varepsilon \, \lambda \, 
\partial_\beta  A_{\rho}
\partial_\alpha A_{\nu}.
\end{eqnarray}
Since $\varepsilon \, \lambda$ is a scalar quantity, the r.h.s. is covariant.
~\newline

Now, we consider some $\tilde{g}$ connecting $g$ in (\ref{tyjyr}) and $1$ as follows:
\begin{eqnarray}\label{rtppo}
\{ \tg(x,t) \}_{0 \le t \le 1}: 
\textrm{
$\tg(x,0) = g(x)$ 
~and~ 
$\tilde{g}(x,1) = 1$}.
\end{eqnarray}
By this, we can see $g(x)$ as $\tg(x,t)$, generally; 
therefore, we consider $\Gamma[g,A]_{\rm P} $ as $\Gamma[\tg,A]_{\rm P} $ in what follows.
%---
Now, we present the transformations for $t$ as
\begin{subequations}\label{wbsrj}
\begin{align}
\label{wbsrj1}
\tilde{g}(x,t+dt) 
=& \,\, \underbrace{\tg(x,t+dt)\,\tg^{-1}(x,t)}_{\equiv \, U(x,t)} \, \tilde{g}(x,t)
= \tilde{g}(x,t) + \frac{\partial \tg(x,t)}{\partial t}dt +{\cal O}(dt^2), \\*
\label{wbsrj2}
A_\mu^{U(x,t+dt)} 
=& \,\, A_\mu^{U(x,t)} + iU(x,t) \partial_\mu U^{-1}(x,t),
\end{align}
\end{subequations}
where 
$\{ U(x,t) \}_{0 \le t \le t_{\rm T}(x)}:$ 
$U(x,0) = 1$ and $U(x,t_{\rm T}(x)) = g^{-1}(x)$.
$A_\mu^{U(x,t)}$ is the $A_\mu$ at the point $\tilde{g}=\tilde{g}(x,t)$. 
We denote $U(x,t)$ just as $U(t)$ in what follows.
Then, since the transformations in (\ref{yjpr}) can be considered as the infinitesimal variations in (\ref{wbsrj}), 
we can present the gauge transformations in (\ref{mtth}) with the parametrization with $t$. 
Therefore, by taking $\varepsilon\lambda$ as
\begin{eqnarray}\label{mythm}
\varepsilon\lambda=\Delta(x) dt,
\end{eqnarray}
we can present $\delta_\varepsilon$ in (\ref{mtth}) 
as $\delta_\varepsilon = dt \,\partial/\partial t$,
where
$\Delta(x)$ is an adjustment factor,
$dt$ is constant both for $t$ and $x$, while
$\varepsilon\lambda$ depends on $x$, but constant for $t$, as oen variable. 
%---
Based on these, we can present (\ref{mtth}) as
\begin{eqnarray}\label{peter}
dt \,\frac{\partial}{\partial t} \Gamma_{\rm P}[\tilde{g}(t),A^{U(t)}]
= - \frac{ie^2\epsilon^{\beta \rho \alpha \nu}}{4\pi^2}
\int d^4x \,\Delta \,dt\, 
\partial_\beta  A_{\rho}^{U(t)}
\partial_\alpha A_{\nu}^{U(t)}.
\end{eqnarray}
Therefore, 
we can obtain the following result:
\begin{eqnarray}\label{nriu}
\Gamma_{\rm P}[1,A^{g^{-1}}]-\Gamma_{\rm P}[g,A]  
= -\frac{ie^2}{24\pi^2} \int d^4x \,\Delta
\int_0^1 dt \, 
\partial_\beta  A_{\rho}^{U(t)}
\partial_\alpha A_{\nu}^{U(t)}.
\end{eqnarray}
However, 
in the one above, 
we can find that 
the overall factor of $\Gamma_{\rm P}$ 
will differ up to $\Delta$, 
which is a problem.
We will remedy this point.

For this purpose, we retake (\ref{rtppo}) as follows:
\begin{eqnarray}\label{rlvio}
\{ \tg(x,t) \}_{0 \le t \le t_{\rm T}(x)}:&& \hspace{-6.5mm}
\textrm{
$\tg(x,0) \equiv g(x)$ 
~and~ 
$\tilde{g}(x,t_{\rm T}(x)) \equiv 1$},\\
&& \hspace{-5mm}
\textrm{$t_{\rm T}(x)=1$ when $\Delta(x)=1$,}
\nonumber
\end{eqnarray}
where $t_{\rm T}$ is constant for $t$, although differs at each $x$. 
Then, making use of the two properties:
${\bf 1)}$\, The parametrization of $t$ that satisfies (\ref{rlvio}) can be performed totally freely,  
${\bf 2)}$\, $t_{\rm T}$ can differ for each $x$, although it is constant for $t$;
we can see that we can take $t_{\rm T}(x)$  in such a way that  
$\Delta(x)$ can be canceled, for every $x$
(why this is always possible can be understood from an elementary discussion, 
however, this becomes lengthy if described in full; 
therefore, we omit to describe this). 
Therefore, we can obtain the following equation instead of (\ref{nriu}):
\begin{eqnarray}\label{toki}
\Gamma_{\rm P}[1,A^{g^{-1}}]-\Gamma_{\rm P}[g,A]  
= -\frac{ie^2}{24\pi^2} \int d^4x \,  \int_0^{\bar{t}_{\rm T}(x)} dt \, 
\partial_\beta  A_{\rho}^{U(t)}
\partial_\alpha A_{\nu}^{U(t)},
\end{eqnarray}
where $\bar{t}_{\rm T}$ presents the suitably taken $t_{\rm T}$ in such a way that $\Delta$ becomes 1 at every $x$.
$\Gamma_{\rm P}[1,A^{g^{-1}}]$  is constant for $t$. 
This is because $\Gamma_{\rm P}[1,A^{g^{-1}}]$ is at the upper bound of $t$.
From this, it can be seen that 
$\Gamma_{\rm P}[1,A^{g^{-1}}]$ can be classified to a homogeneous solution, 
and $\Gamma_{\rm P}[g,A]$ is ultimately the special solution, and so-called Wess-Zumino-Witten term. 
\newline

In the discussion up until this point,
we can find that there is a freedom in how to set the overall factor. 
Namely, some function can always appear up to how to take $t_{\rm T}$.
%---
This freedom can be considered as an ambiguity not to be fixed theoretically.
Therefore, we will fix it from the consistency with the observation.

As mentioned in the beginning of this subsection, 
$\Gamma_{\rm P}$ in (\ref{fver}) is analogous to a low-energy effective model 
considered based on the anomalous SU$_\textrm{c}$(3) QCD model 
(at this time, $U(x,t)$ is set to the SU$_\textrm{c}$(3) pseudoscalar boson octet representation).   
%---
The decay-widths 
calculated from that 
are known to be able to agree with the experimental data very well.
%---
At this time, the part to give the overall factor is given as $-\frac{ie^2}{24\pi^2} \int d^4x \cdots$ like (\ref{toki}) 
without any extra function.
From this, we can see it is observationally right 
to take $t_{\rm T}$ so that 
$\Delta$ becomes $1$ for every $x$ in (\ref{nriu}) as well.
%---

The problem concerning this freedom also exists in the case of U(1) gauge theory.
However, in fact, this problem would not be taken up as a problem.
This would be because
the matters discussed in this subsection are usually addressed with the differential form in other discussions; 
therefore, the problems arisen from the integral regions and integral measures are likely to be missed. 
Since these have been explicitly treated in our discussion, 
the problem concerning these has emerged in our discussion.
\newline

The form of (\ref{toki}) is identical with the equation in the case of the U(1) gauge theory. 
Therefore, new effective model would not be obtained
even if we continued the calculation from (\ref{toki}).
However, this is a normal result, because the effective model obtained in this subsection is a low-energy effective model, 
the existence of which links to the expectations for new phenomenology. 
%---
However, no new phenomenology, for which we need new effective model to explain it, 
has not been detected, in practice. 

\vspace{-1.0mm}
%=============================================
\subsection{The future development}
\label{ityrn}
%=============================================  

In the preceding subsections, 
we discussed issues related to our large chiral symmetry breaking.
In this subsection, 
based on that, 
we comment on the possibility of deriving \textit{the anomaly equations of the large gauge symmetries}
as a future development of this study.

In fact, the large gauge symmetries are known 
to be broken~\cite{Strominger:2017zoo}. 
Consequently, the WT identities 
with respect to the BRS transformations 
associated with the large gauge symmetries 
do not hold, and anomalies of these broken BRS transformations arise.
However, neither the anomalies themselves nor the corresponding anomaly equations 
have been derived so far.
This project could make such a derivation possible 
by employing the method in Sec.\,\ref{ypvsb} of this study.

As one of the issues in which anomaly equations play an essential role, 
the derivation of the Hawking radiation 
by the anomaly cancellation method\,\cite{Iso:2006wa}
may be considered.
%-----
In fact,  
in the near-horizon region,
the classical motion of all particles 
is confined to a single direction toward the horizon.
Correspondingly,
in the anomaly cancellation method, 
the conservation laws in this region are constructed 
solely from a single chiral sector.
These conservation laws are anomalous and correspond  to the anomaly equations.  
%-----
It is considered that there is a fundamental difference between
these anomaly equations and the anomaly equations of the large gauge symmetries, 
specifically regarding whether they are anomalous or not from the outset. 
The former are anomalous because they are constructed solely from a single chiral sector, 
whereas the latter are inherently anomalous as a result of some spontaneous symmetry breaking.
%-----

Therefore, it is expected that 
new insights into Hawking radiation may be gained 
from analyzing the anomaly equations of large gauge symmetries 
using the anomaly cancellation method.

\vspace{-1.0mm}
%=============================================
\section{Summary}
\label{iptr}
%=============================================  

In this study, we proposed the large chiral symmetry based on the large U(1) gauge symmetry, 
and obtained the anomaly equation of that.
%---
Subsequently,  
we considered the large BRS-transformed one-loop diagrams 
obtained from the axialization of the large BRS-transformed effective action.
Then, evaluating these one-loop diagrams, 
we showed that 
the anomaly equation of the large chiral symmetry can be derived.
We also showed that it can be derived from the Fujikawa method.
%---
These results  provide an important step in the development of large U(1) gauge symmetry.

The fermionic field in this study was massless. 
If it were massive, 
it would all start off from ${\cal I}^{-}$ and reach  ${\cal I}^{+}$ in Fig.\,\ref{wsdd57}.
In that case, the way to define the soft-charges differs from the massless case in Sec.\,\ref{vhtew1}. 
For this reason, this study did not address the massive Dirac field.  

As a direction for future development, 
we may consider the coupling of the fermion to gravity 
(the fermion in this study is coupling to the U(1) gauge field).
However, the gravity in the asymptotic symmetry is a problem of how the spacetime is curved and anomaly equation is local. 
Therefore, since the anomaly equation is covariant, 
it can  be obtained using only general coordinate transformations 
applied to the anomaly equation given on flat spacetime.
We commented on some possible future development in Sec.\,\ref{ityrn}.

\appendix

\vspace{-2.0mm}
%=============================================
\section{The derivation of (\ref{resyu})}
\label{app:ttih}
%=============================================

We first examine the variation of the action 
for the following general coordinate and  field transformations:
\vspace{-4.0mm}
\begin{eqnarray}\label{evajr}
x^\mu \to  x'{}^\mu       =  x^\mu + \delta x^\mu, \quad
\phi_A(x) \to \phi'_A(x') =  \phi_A(x)+ \delta \phi_A(x),
\end{eqnarray}
where $\phi_A(x)$ represent the fields as defined in Sec.\,\ref{app:ssdik}. 
$\delta x^\mu$ and $\delta \phi_A$ are assumed to be the quantities of the liner order. 
The variation of the action under (\ref{evajr}) can be computed as follows:
\vspace{-1.0mm}
\begin{eqnarray}\label{ahet}
\delta I 
\!\!\! &=& \!\!\! 
 \int_{\Omega'} d^4x' \, L(\phi'_A(x'),\phi'_{A,\mu}(x'))
-\int_\Omega    d^4x  \, L(\phi_A(x),\phi_{A,\mu}(x)) 
\nonumber\\* 
%---
\!\!\! &=& \!\!\! 
\int_\Omega d^4x \,
\Big(
 \frac{\partial L}{\partial \phi_A} \bar{\delta} \phi_A
+\frac{\partial L}{\partial \phi_{A,\mu}}\partial_\mu ( \bar{\delta} \phi_A )
+\partial_\mu(L\,\delta x^\mu)
\Big),
\end{eqnarray}
where 
$d^4x' = \left\vert \frac{\partial x'{}^\mu}{\partial x^\nu} \right\vert d^4x
       = (1+\frac{\partial \delta x^\mu}{\partial x^\mu})\,d^4x$ and 
$
\phi'_{A,\mu}(x')
=\frac{\partial x^\nu}{\partial x'^\mu}\,\frac{\partial }{\partial x^\nu} \phi'_A(x')
=
\partial_\mu\phi_A(x)+\partial_\mu	(\delta \phi_A(x))
-(\partial_\mu \delta x^\nu) \,\partial_\nu \phi_A(x)
$. 
$\bar{\delta} \phi_A(x)$ represents the Lie derivative for $\phi_A$, 
which is defined as $\bar{\delta} \phi_A(x=a) \equiv \phi'_A(x'=a)-\phi_A(x=a)$, 
and $\bar{\delta} \phi_A$ can be written as
$
\bar{\delta} \phi_A
=
\delta \phi_A
-\partial_\mu\phi_A\,\delta x^\mu
$. Finally,
$\delta I$ can be given as
\vspace{-1.0mm}
\begin{eqnarray}\label{esre}
\delta I
=
\int_\Omega d^4x\, 
\Big(
 [L]^A\bar{\delta} \,\phi_A
+\partial_\mu {\cal J}^\mu
\Big), \quad
{\cal J}^\mu \equiv \frac{\partial L}{\partial \phi_{A,\mu}} \delta \phi_A-T^\mu{}_\nu \,\delta x^\nu,
\end{eqnarray}
where $[L]^A$ and $T^\mu{}_\nu$  are defined in (\ref{wevawr}) and (\ref{vrtyz}), respectively. 

Applying the local transformation (\ref{gdwr}) to (\ref{esre}), 
\vspace{-1.0mm}
\begin{eqnarray}\label{saseo}
\delta I\vert_{\rm (\ref{gdwr})}
=
\int_\Omega d^4x \,
\Big\{
\lambda^r 
\Big(
[L]^A(M_{r,A}-\phi_{A,\mu}X^\mu{}_r)
-\partial_\mu([L]^A N_r{}^\mu{}_{,A})
\Big)
+
\partial_\mu \Big(
{\cal J}^\mu
+[L]^A N_r{}^\mu{}_{,A} \, \lambda^r 
\Big)
\Big\},
\end{eqnarray}
where $J_r^\mu$ is defined in (\ref{vrtyz}). 
Writing out $\lambda^r$ included in $\delta \phi_A$ and $\delta x^\nu$ with  (\ref{gdwr}) in the second part, 
\vspace{-1.0mm}
\begin{eqnarray}\label{zsts}
\partial_\mu (B^\mu{}_r \, \lambda^r +C^{\mu\nu}{}_r \, \lambda^r{}_{,\nu}) 
=
  \partial_\mu B^\mu{}_r \, \lambda^r
+ (B^\mu{}_r+\partial_\nu C^{\nu\mu}{}_r) \, \lambda^r{}_{,\mu}
+ C^{\mu\nu}{}_r  \, \lambda^r{}_{,\mu\nu},
\end{eqnarray}
and (\ref{resyu}) can be obtained.
(On the other hand, it can be seen from the first part that 
there are $n$ equations ($r=1,\cdots,n$) 
among $N$ EL equations for $\phi_A$ ($A=1,\cdots,N$). 
This means that 
the EL equations are $N-n$ missing for determining $N$ $\phi_A$, 
which leads to the issue of a constrained system.)  

\vspace{-2.0mm}
%=============================================
\section{The formulas used in Sec.\,\ref{ypvsb}}
\label{app:orrew}
%=============================================
\vspace{-3.0mm}
\begin{subequations}
\begin{align}
\label{mrrgy1}
\frac{1}{ABC} 
 =& \,\, 
2 \int_0^1 y \, dy \int_0^1 dx \,\frac{1}{( Ayx +By(1-x) +C(1-y) )^3},%\\* 
\\*
%=====
\label{mrrgy5}
\int \! \frac{d^nk}{i(2\pi)^n}\frac{k^\mu k^\nu}{{\cal B}^\alpha}
=& \,\, 
\frac{1}{{\cal D}}\,
\Big( 
\Gamma(\bar{\alpha})   \frac{p^\mu p^\nu}{{\cal C}^{\bar{\alpha}}}
-\Gamma(\bar{\alpha}-1) \frac{g^{\mu\nu}}{2{\cal C}^{\bar{\alpha}-1}}
\Big),
\\*
%=====
\label{mrrgy3}
\int \! \frac{d^nk}{i(2\pi)^n}\frac{k^\mu k^\nu k^\rho}{{\cal B}^\alpha}
=& \,\, 
\frac{1}{{\cal D}}\,
\Big(  
\Gamma(\bar{\alpha})    \frac{p^\mu p^\nu p^\rho}{{\cal C}^{\bar{\alpha}}}
-\Gamma(\bar{\alpha}-1) \frac{g^{(\mu \nu}p^{\rho)}}{2{\cal C}^{\bar{\alpha}-1}}
\Big), 
\\*
%=====
\label{mrrgy4} 
\int \! \frac{d^nk}{i(2\pi)^n}\frac{k^\mu k^\nu k^\rho k^\sigma}{{\cal B}^\alpha}
=& \,\, 
\frac{1}{{\cal D}}\,
\Big( 
\Gamma(\bar{\alpha})    \frac{p^\mu p^\nu p^\rho p^\sigma}{{\cal C}^{\bar{\alpha}}}
-\Gamma(\bar{\alpha}-1) \frac{g^{(\mu \nu}p^\rho p^{\sigma)}}{2{\cal C}^{\bar{\alpha}-1}}
+\Gamma(\bar{\alpha}-2) \frac{g^{(\mu \nu}g^{\rho\sigma)}}{4{\cal C}^{\bar{\alpha}-2}}
\Big), 
\end{align}  
\end{subequations}
where $\eta \equiv n/2$, 
${\cal B} \equiv m^2+2 k \cdot p-k^2$, 
${\cal C} \equiv m^2+p^2$, 
${\cal D} \equiv (4\pi)^\eta \,\Gamma(\alpha)$, 
$\bar{\alpha}\equiv \alpha-\eta$ and 
\vspace{-1.0mm}
\begin{eqnarray}
g^{(\mu \nu}p^{\rho)} 
\!\!\! &=&\!\!\! 
 g^{\mu \nu} p^{\rho}
+g^{\nu\rho}p^{\mu}
+g^{\rho\mu}p^{\nu},
\nonumber\\*
g^{(\mu \nu}     p^\rho p^{\sigma)} 
\!\!\! &=&\!\!\!  
 g^{\mu \nu}     p^\rho p^\sigma
+g^{\mu \rho}    p^\nu  p^\sigma
+g^{\mu \sigma}  p^\nu  p^\rho 
+g^{\nu \rho}    p^\mu  p^\sigma
+g^{\nu \sigma}  p^\mu  p^\rho 
+g^{\rho \sigma} p^\mu  p^\nu, 
\nonumber\\*
g^{(\mu \nu}g^{\rho\sigma)}        
\!\!\! &=&\!\!\!  
  g^{\mu \nu}g^{\rho\sigma} 
+ g^{\mu \rho}g^{\nu \sigma} 
+ g^{\mu \sigma}g^{\nu \rho}. 
\nonumber
\end{eqnarray}
(\ref{mrrgy1}) is a Feynman parameter formula. 
(\ref{mrrgy5})-(\ref{mrrgy4}) can be known from Sec.\,A.4 in \cite{Peskin:1995ev}. 

\vspace{-2.0mm}
%=============================================
\section{The Wess-Zumino consistency condition used in Sec.\,\ref{psent1}}
\label{app:stod}
%=============================================

In this Appendix, we first derive the Ward-Takahashi (WT) identity, 
from which we derive the Wess-Zumino consistency condition (WZ condition) for use in the discussion in Sec.\,\ref{psent1}.

\vspace{-1.0mm}
%=============================================
\subsection{The WT identity}
\label{app:stod1}
%=============================================

For the action $S_0$ given in (\ref{erer}), 
We consider the following general liner gauge-fixing condition:
\begin{eqnarray}\label{bieer}
F = 
\partial^\mu A_\mu 
+f \psi
+\alpha B /2 
+ w, \quad \textrm{$f$, $\alpha$ and $w$: some constants},
\end{eqnarray}
At this time, following the BRST formalism, 
the gauge-fixing + FP term in the action is given as follows:
\begin{eqnarray}\label{tbnr}
S_{\textrm{GF+FP}}
=
-i\int d^4x \bdelta_{\textrm{B}} (\bar{c}F)
=
\int d^4x 
{\cal L}_{\textrm{GF+FP}}, \quad
{\cal L}_{\textrm{GF+FP}}
=
BF
+i\bar{c}(e^{-1}\partial^\mu \p_\mu c+ie f c \psi 
),
\end{eqnarray}
where $e$ is that in (\ref{vwrer}), and the BRS transformation of each field is  
\begin{eqnarray}\label{usrvt}
\bdelta_{\textrm{B}}A_\mu=e^{-1}\p_\mu c, \quad
\bdelta_{\textrm{B}}\psi=ie c \psi, \quad
\bdelta_{\textrm{B}}B=\bdelta_{\textrm{B}}c=0, \quad
\bdelta_{\textrm{B}}\bar{c}=iB,
\end{eqnarray}
where the BRS transformation is given by
$\bdelta_{\textrm{B}} \Phi_I=[iQ_{\textrm{B}},\Phi_I]$  
($\Phi_I$ means each of these all fields, 
$Q_{\textrm{B}}$ is the BRS charge, 
and $[\ast,\ast]$ is the commutator. If $\Phi_I$ is Grassmann-odd, 
it is given by the  the anti-commutator $\{\ast,\ast\}$).
$\bdelta_{\textrm{B}} A_\mu$ and $\bdelta_{\textrm{B}} \psi$ can be read from (\ref{fytou}).
%---
Since the gauge field in this study is U(1), 
we omit terms associated with the Yang-Mills sector.
$\psi_i$ and $B$ are the Dirac field  and the NL field, respectively.
%---
Note that the action with $S_{\textrm{GF+FP}}$ in (\ref{tbnr}) is invariant under the BRS symmetry, 
while it is not invariant under the gauge transformations.
 
Now, we consider the term associated with all external fields: 
$A_\mu$, $\psi_i$, $B$, $c$, $\bar{c}$,  
and the BRS transformations of these fields as
\begin{eqnarray}\label{ureev}
S_{\textrm{ext}}[J,K]
\!\!\! &=& \!\!\!
\int d^4x(
J^\mu A_\mu
+J^i \psi_i
+\bar{J}_{c}c
+J_{\bar{c}}\bar{c}
+J_BB
+K^\mu e^{-1}\partial_\mu c
+ K i e c \psi),\nonumber\\*[-1.0mm]
\!\!\! &\equiv& \!\!\!
J^\mu \cdot A_\mu
+J^i \cdot \psi_i
+\bar{J}_{c}c
+J_{\bar{c}}\cdot \bar{c}
+J_B \cdot B
+K^\mu \cdot e^{-1}\partial_\mu c
+K \cdot i e c \psi, 
\quad \cdot \equiv \int dx^4,
\end{eqnarray}
where $J^\mu$ is $c$-numbers, and 
$J^i$, $\bar{J}_{c}$, $J_{\bar{c}}$,  $K^\mu$ and $K^i$ are Grassmann numbers.
%---
Then, the physical states are defined as BRS invariant: 
$Q_{\textrm{B}}\vert {\rm phys} \rangle=0$, 
and the vacuum is a physical state, 
therefore, $Q_{\textrm{B}}\vert 0 \rangle=0$.
Therefore, the WT identity
$0 =
\langle 0 \vert \{iQ_{\textrm{B}},{\rm T}({\cal O}(x^1){\cal O}(x^2)\cdots{\cal O}(x^n)) \} \vert 0 \rangle
$ identically hold, 
where ${\rm T}$ denotes the time-ordered product 
and ${\cal O}(x^k)$ is some operator (Grassmann-even).
Supposing $n=1$ and considering ${\cal O}(x^1)=\exp iS_{\textrm{ext}}$,
we can obtain the following the WT identity: 
\begin{eqnarray}\label{cwye}
0 
\!\!\! &=& \!\!\!
\langle 0 \vert [iQ_{\textrm{B}},{\rm T} \exp iS_{\textrm{ext}}] \vert 0 \rangle
\nonumber \\*[0.5mm]
%---
\!\!\! &=& \!\!\! 
i \langle 0 \vert {\rm T} (
J^\mu \cdot e^{-1}\p_\mu c
-J \cdot iec\psi
-J_{\bar{c}} \cdot iB)
\exp i S_{\textrm{ext}}\vert 0 \rangle,
\end{eqnarray}
where
$\exp i S_{\textrm{ext}}$ denotes $\exp [i S_{\textrm{ext}}]$.
In the calculation above, 
the relations:
$
\{iQ_{\textrm{B}},c\}
=[iQ_{\textrm{B}},B]
=\{iQ_{\textrm{B}},e^{-1}\p_\mu c\}
=[iQ_{\textrm{B}},iec\psi] =0
$, which follow from the nilpotency of the BRST charge, were used.

Here, we define 
the generating functional for connected Green's functions $W$  
and the generating functional for 1PI vertex functions $\Gamma$ as
\begin{subequations}
\begin{align}  
\label{brtn1}
\exp iW[J,K]
\equiv& \,\,
\langle 0 \vert
{\rm T} \exp iS_{\textrm{ext}}
\vert 0 \rangle, \\*
%-----
\label{brtn2}
\Gamma[\Phi,K] 
\equiv& \,\,
W[J,K]-J^I \cdot \Phi_I, \quad 
\Phi_I = \frac{\delta W}{\delta J^I}
=\frac
{ \langle 0 \vert  {\rm T} \Phi_I \exp i S_{\textrm{ext}} \vert 0 \rangle}
{\langle 0 \vert {\rm T} \exp i S_{\textrm{ext}} \vert 0 \rangle}.
\end{align}  
\end{subequations}
Therefore, the following equations hold:
\begin{eqnarray} \label{tjjewe}
\frac{\delta \Gamma}{\delta \Phi_I} = -(-)^{\vert I \vert} J^I, \quad
\frac{\delta W }{\delta K^\mu}   = \frac{\delta  \Gamma}{\delta K^\mu}, \quad
\frac{\delta W }{\delta K}   = \frac{\delta  \Gamma}{\delta K},
\end{eqnarray}
where $\vert I \vert$ is the statistical index of $J^I$.
Then, the WT identity (\ref{cwye}) can be presented  as 
\begin{eqnarray}\label{rlbwe}
(
  J^\mu \cdot \frac{\delta}{\delta K^\mu}
- J \cdot \frac{\delta}{\delta K}
-iJ_{\bar{c}}\frac{\delta}{\delta J_B}
)W=0.
\end{eqnarray}
This can be given as 
\begin{eqnarray}\label{dvndl}
   \frac{\delta \Gamma}{\delta A_\mu}   \cdot \frac{\delta \Gamma}{\delta K^\mu}
+  \frac{\delta \Gamma}{\delta \psi}  \cdot \frac{\delta \Gamma}{\delta K}
+i \frac{\delta \Gamma}{\delta \bar{c}} \cdot B=0,
\end{eqnarray}
where $\frac{\delta }{\delta J_B}W[J,K] =\frac{\delta}{\delta J_B}(\Gamma[\Phi,K] +J^I \cdot \Phi_I)=B$.

\vspace{-1.0mm}
%=============================================
\subsection{The brief expression of WT identity}
\label{app:stod3}
%=============================================

We have obtained the WT identity in (\ref{dvndl}).
In this subsection, we give a brief expression of it.
\newline

From the stationary condition for $\bar{c}$ in the system given 
by the action with $S_{\rm GF+FP}$ and $S_{\rm ext}$, 
the field equation of $c$ can be obtained as $i\p_\mu \p^\mu c +J_{\bar{c}}=0$,
which can be given as
\begin{eqnarray}\label{wreyj}
e\partial^\mu \frac{\delta \Gamma}{\delta K^\mu}+i \frac{\delta \Gamma}{\delta \bar{c}}=0.
\end{eqnarray}
From this, it can be seen that  
the $\bar{c}$-dependence appears via 
$\tilde{K}^\mu \equiv K^\mu+ie\bar{c}\p^\mu$ in $\Gamma$, 
and we can retake the independent variables of $\Gamma$ as 
\begin{eqnarray}\label{rebf}
\Gamma[A_\mu, \psi, c,\bar{c},K^\mu,K] \to 
\Gamma[A_\mu, \psi, c,\tilde{K}^\mu,K]
\end{eqnarray}
by rewriting each $K^\mu$ in $\Gamma$ 
as $\tilde{K}^\mu-ie\bar{c}\p^\mu$\,\footnote{
%======================
%===== FOOT NOTE ======
%======================
As can be seen from (\ref{brtn2}), $\Gamma$ is originally given as 
$\Gamma=\Gamma[A_\mu, \psi_i, c, \bar{c} ,K^\mu, K^i]$.
We retake its independent variables as
$\Gamma=\Gamma[A_\mu, \psi_i, c, \bar{c} ,\tilde{K}^\mu, K^i]$. 
Then, from  (\ref{wreyj}) for this $\Gamma$, 
${\delta \Gamma}/{\delta \bar{c}}=0$ can be derived as follows:
\begin{eqnarray}\label{irge}
0=e\partial^\mu \frac{\delta \Gamma}{\delta K^\mu}
+i \frac{\delta \Gamma}{\delta \bar{c}}
=
e\partial^\mu (
\frac{\delta \tilde{K}^\nu}{\delta K^\mu} 
\frac{\delta \Gamma}{\delta \tilde{K}^\nu}
)
+i (
\frac{\delta \bar{c}}{\delta \bar{c}}
\frac{\delta \Gamma}{\delta \bar{c}}
+
\frac{\delta \tilde{K}^\nu}{\delta \bar{c}}
\frac{\delta \Gamma}{\delta \tilde{K}^\nu}
)
=
\hspace{-3mm}
\underbrace{(e\partial^\mu \frac{\delta \tilde{K}^\nu}{\delta K^\mu}
+i\frac{ \delta \tilde{K}^\nu}{\delta \bar{c}})}_{\hspace{4mm}\textrm{$=0$ as $\tilde{K}^\mu \equiv K^\mu+ie\bar{c}\p^\mu$}} 
\hspace{-3mm}
\frac{\delta \Gamma}{\delta \tilde{K}^\nu}
+i \frac{\delta \bar{c}}{\delta \bar{c}}\frac{\delta \Gamma}{\delta \bar{c}}
=i \frac{\delta \Gamma}{\delta \bar{c}}.
\end{eqnarray}
%======================
%======================
%======================
}.  
%---
Based on the point above, (\ref{dvndl}) can be equally given as follows:  
\begin{eqnarray}\label{vaewj}
   \frac{\delta \Gamma}{\delta A_\mu}   \cdot \frac{\delta \Gamma}{\delta \tilde{K}^\mu}
+  \frac{\delta \Gamma}{\delta \psi}  \cdot \frac{\delta \Gamma}{\delta K}=0.
\end{eqnarray}

Here, we introduce the operator $\ast$ as follows:
\begin{eqnarray}\label{sjyn}
F \ast G \equiv 
   \frac{\delta F}{\delta A_\mu}  \cdot \frac{\delta G}{\delta \tilde{K}^\mu} 
+  \frac{\delta F}{\delta \psi} \cdot \frac{\delta G}{\delta K}
+(-)^{\vert F \vert}(
   \frac{\delta F}{\delta A_\mu}  \cdot \frac{\delta G}{\delta \tilde{K}^\mu} 
+  \frac{\delta F}{\delta \psi} \cdot \frac{\delta G}{\delta K}).
\end{eqnarray}
For this operation with $\ast$, the following formulas hold for the generally $F$, $G$ and $H$:
\begin{subequations}
\begin{align} 
\label{svti1}
& 
F \ast G = 
-(-)^{(\vert F \vert +1)(\vert G \vert +1)}
G \ast F,\\*
\label{svti2}
&
F \ast (G \ast H) 
+ (-)^{(\vert F \vert +1)(\vert G \vert +\vert H \vert+2)}G \ast (H \ast F)
+ (-)^{(\vert H \vert +1)(\vert F \vert +\vert G \vert+2)}H \ast (F \ast G)=0.
\end{align} 
\end{subequations}
Using this $\ast$, we can briefly present the WT identity (\ref{vaewj}) as follows:
\begin{eqnarray}\label{vvbb}
\Gamma \ast \Gamma =0.
\end{eqnarray}

\vspace{-6.0mm}
%=============================================
\subsection{The WZ condition}
\label{app:stod2}
%=============================================

In this subsection, we derive the WZ condition. 
To this end, we begin with an effective action 
given by the loop-expansion to the $n$-loop order as
\begin{eqnarray}\label{urdvv}
(\Gamma)_n 
\equiv 
\Gamma^{(0)}
+\hbar \Gamma^{(1)}
+\cdots
+\hbar^n \Gamma^{(n)}.
\end{eqnarray}
First, since $\Gamma^{(0)}$ is always given solely by the classical Lagrangian, 
$\Gamma^{(0)}$ exactly satisfies the WT identity as $\Gamma^{(0)} \ast \Gamma^{(0)}=0$.
%---
Now, we suppose that 
some regularization of the UV-divergences has been implemented in this $(\Gamma)_n$.
If this regularization did not break the BRS symmetry, 
$(\Gamma)_n$ would satisfy the WT identity up to $\hbar^n$-order as follows:
\begin{eqnarray}\label{ervds}
(\Gamma)_n \ast (\Gamma)_n = {\cal O}(\hbar^{n+1}).
\end{eqnarray}
The theory is said to be renormalizable if this holds for arbitrary $n$. 
However, the BRS symmetry may be broken 
by the regularization in the $\hbar^{n+1}$-order loops.
At this time, while the WT identity is preserved up to $\hbar^n$-order as in (\ref{ervds}), 
it is broken at $\hbar^{n+1}$-order as
\begin{eqnarray}\label{svrbe}
(\Gamma)_{n+1}  \ast (\Gamma)_{n+1}  = \hbar^{n+1} 2 \varDelta+ {\cal O}(\hbar^{n+2}),
\end{eqnarray}
where $n+1=1,2,\cdots$ and
\begin{itemize}
\item
The equation above shows that 
the BRS symmetry of the system is broken at the $\hbar^{n+1}$-order. 
Its cause is the regularization of the UV-divergence.
This is because the BRS symmetry would always be preserved and $\varDelta=0$, 
if any modifications were not implemented on the theory
and the modification of the theory is brought about 
by the regularization of the UV-divergence.
\end{itemize}

However, there is the case that $\varDelta$ can also be expressed in the following way:
\begin{eqnarray}\label{rnrte}
\varDelta = (\Gamma)_n  \ast Y
\end{eqnarray}
using some function $Y$.
In this case, we newly retake the following one as the effective action:
\begin{eqnarray}\label{oite}
(\Gamma)_{n+1} - \hbar^{n+1} Y
= 
(\Gamma)_n 
+ \hbar^{n+1}
\hspace{0mm}
\underbrace{(\Gamma^{(n+1)}-Y)}_{\hspace{0mm}\textrm{new $\Gamma^{(n+1)}$}}
\hspace{0mm}
\equiv (\tilde{\Gamma})_{n+1},
\end{eqnarray}
where its $(\Gamma)_n$ and $(\Gamma)_{n+1}$ are those in (\ref{ervds}) and (\ref{svrbe}). 
Then, by using (\ref{rnrte}) in this new $(\tilde{\Gamma})_{n+1}$,
the broken WT identity (\ref{svrbe}) can be remedied to be preserved to the $\hbar^{n+1}$-order as follows:
\begin{eqnarray}\label{erbrt}
(\tilde{\Gamma})_{n+1} \ast (\tilde{\Gamma})_{n+1}
=\hbar^{n+1} \{2 \varDelta-((\Gamma)_n \ast Y+Y \ast (\Gamma)_n )\}+{\cal O}(\hbar^{n+2})
={\cal O}(\hbar^{n+2}).
\end{eqnarray}

However, there is the case that we cannot express $\varDelta$ 
in the way of (\ref{rnrte}) 
by any means.
In this case, we have to accept the fact that 
the BRS symmetry of the system is broken at the $\hbar^{n+1}$-order 
in exchange for the regularization. 
%---
At this time, by operating $(\Gamma)_{n+1} \ast$ to both sides of (\ref{svrbe}),
we can obtain an equation for $\varDelta$ as
$(\Gamma)_{n+1} \ast \varDelta +{\cal O}(\hbar)=0$,
where we have used (\ref{svti2}) and $n+1=1,2,\cdots$. 
Since this is the equation to  the $\hbar^1$-order, 
we can obtain the following equation from this:
\begin{eqnarray}\label{briu0}
\Gamma^{(0)}  \ast \varDelta = \bdelta_{\textrm{B}} \varDelta = {\cal O}(\hbar).
\end{eqnarray}
where the operator $\Gamma^{(0)} \ast$ is equal to the BRS transformation 
(which we can see from (\ref{sjyn})), and
\begin{itemize}
\item
Under (\ref{svrbe}), we noted the cause of the appearance of $\varDelta$: 
it is a by-product of the regularization 
which cannot be eliminated 
by any modification of the regularization scheme.
%---
However, $\varDelta$ itself is merely a solution 
and independent of the specific regularization employed. 
Therefore, the form of $\varDelta$ is universal across different regularization schemes 
and merely presents a possible form that can appear in (\ref{svrbe}).
%---
One point we can note from this is that
whether $\varDelta$ always appears or not in other regularizations is a separate issue,
even if it appeared in a regularization. 

Regarding this, 
it can be concluded that $\varDelta$ necessarily appears
if it appears in one regularization.
%---
This is because 
employing another regularization can be regarded 
as a modification of the regularization scheme 
in use at that time. 
%---
However, $\varDelta$ we are dealing with here is precisely that 
which cannot be eliminated by any modification, 
based on the supposition around (\ref{rnrte}).
%---
Therefore, if $\varDelta$ appears in one regularization,
it will appear in other regularizations as well.
\end{itemize}

\vspace{-4.0mm}
%=============================================
\subsection{Brief comment on the solution of the WZ condition}
\label{app:stod3}
%=============================================

In this subsection, we briefly comment on how to solve it and on its solution.
First, we can see that 
$\varDelta$ satisfying the WZ condition can be given as follows:
\begin{eqnarray}\label{grrsd}
   \varDelta = \int d^4x \,\mathfrak{a}, \quad
\mathfrak{a} = \mathfrak{a}^{\textrm{(trivial)}}+\mathfrak{a}^{\textrm{(non-trivial)}}+{\cal O}(\hbar),
\end{eqnarray}
where 
\begin{itemize}
\item
The reason $\varDelta$ is expressed as a spacetime integral is that  
the WZ condition (\ref{briu0}) can be satisfied locally   
even if its spacetime integral is omitted
(which is denoted with the notation in (\ref{ureev})).

If $\mathfrak{a}=\bdelta_{\textrm{B}} (\cdots)+{\cal O}(\hbar)$ 
(any quantities can be used for ``$\cdots$''), 
$\varDelta$ is a solution of (\ref{briu0}) because of $\bdelta_{\textrm{B}}^2=0$.
In (\ref{grrsd}), $\mathfrak{a}^{\textrm{(trivial)}}$ denotes this $\mathfrak{a}$.

\item 
On the other hand, if either:
$\bdelta_{\textrm{B}}\mathfrak{a}={\cal O}(\hbar)$
or $\bdelta_{\textrm{B}} \mathfrak{a}=\partial_\mu(\cdots)^\mu+{\cal O}(\hbar)$, 
$\varDelta$ is a solution of (\ref{briu0}).
In (\ref{grrsd}), $\mathfrak{a}^{\textrm{(non-trivial)}}$ denotes this case.
\end{itemize}
Therefore, the task to obtain the solution reduces to obtaining $\mathfrak{a}^{\textrm{(non-trivial)}}$.

To this end, we first need to take into account the following fact, 
which was obtained in the discussion of renormalization theory:
\begin{eqnarray}\label{vrrtj}
\begin{array}{l}
\textrm{$\varDelta$ is a quantity with $N_{\rm FP}=1$ and the dimension less than or equal to $5$ with mass dimension.} 
\end{array}
\end{eqnarray}
With this in mind, 
one way to obtain $\mathfrak{a}^{\textrm{(non-trivial)}}$ is 
to write down all terms allowed by (\ref{vrrtj}), 
then identify those corresponding to $\mathfrak{a}^{\textrm{(non-trivial)}}$. 
At this time, 
\begin{itemize}
\item
In the case where the gauge group associated with $\varDelta$ is simple,
we can identify $\mathfrak{a}$ 
corresponding to 
$\bdelta_{\textrm{B}} \mathfrak{a}=\partial_\mu(\cdots)^\mu+{\cal O}(\hbar)$ 
up to an overall factor 
(from the discussion in the BRST formalism, 
it follows for a simple group that 
$\mathfrak{a}$ satisfying
$\bdelta_{\textrm{B}}\mathfrak{a}={\cal O}(\hbar)$ 
is excluded under (\ref{vrrtj})). 
This satisfies: $\bdelta_{\textrm{B}} (\Gamma)_n \propto \mathfrak{a}$.
This establishes the link between the solution of the WZ condition and the anomaly.
%---

\item
On the other hand, 
in the case where the gauge group is U(1), 
the solution of $\mathfrak{a}$ obtained in the above simple group case can be directly applied
by interpreting the gauge field in that analysis as the U(1) gauge field.
%---
However, 
due to the commutativity of the U(1) gauge field, 
several terms in $\mathfrak{a}$ vanish. 
In addition, 
while $\bdelta_{\textrm{B}} \mathfrak{a}$ can be presented as $\partial_\mu(\cdots)^\mu+{\cal O}(\hbar)$, 
it identically vanishes (see (\ref{gvdy}) for the concrete example of this). 
%---
At this time, $\mathfrak{a}$ also satisfies: $\bdelta_{\textrm{B}} (\Gamma)_n \propto \mathfrak{a}$. 
However, some kind of axilization has been performed in $\bdelta_{\textrm{B}} (\Gamma)_n$ 
(what we have done for this is (\ref{iukne})). 
\end{itemize}

\vspace{-5.0mm}
%=============================================

\end{document}